\documentclass[physrev,reprint,bibnotes,citeautoscript]{revtex4-2}  

\usepackage[T1]{fontenc}
\usepackage[utf8]{inputenc}
\usepackage[australian]{babel}
\usepackage[a4paper,centering,hmargin=1.75cm,vmargin=2cm]{geometry} 
\usepackage{amsmath,amssymb,graphicx,bm,microtype}
\usepackage[dvipsnames]{xcolor}
\usepackage[colorlinks,allcolors=blue!50!black]{hyperref}
\usepackage[all]{hypcap}
\usepackage{cleveref}
\usepackage{braket}
\usepackage{siunitx} 

\newcommand{\abs}[1]{\lvert #1 \rvert}

\begin{document}

\title{Jumping kinetic Monte Carlo: Fast and accurate simulations of partially delocalised charge transport in organic semiconductors}

\author{Jacob T. Willson}
\affiliation{School of Chemistry, University of Sydney, NSW 2006, Australia}

\author{William Liu}
\affiliation{School of Chemistry, University of Sydney, NSW 2006, Australia}

\author{Daniel Balzer}
\affiliation{School of Chemistry, University of Sydney, NSW 2006, Australia}

\author{Ivan Kassal}
\email[Email: ]{ivan.kassal@sydney.edu.au}
\affiliation{School of Chemistry, University of Sydney, NSW 2006, Australia}


\begin{abstract}
Developing devices using disordered organic semiconductors requires accurate and practical models of charge transport. In these materials, charge transport occurs through partially delocalised states in an intermediate regime between localised hopping and delocalised band conduction. Partial delocalisation can increase mobilities by orders of magnitude over conventional hopping, making it important for materials and device design. Although delocalisation, disorder, and polaron formation can be described using delocalised kinetic Monte Carlo (dKMC), it is a computationally expensive method. Here, we develop jumping kinetic Monte Carlo (jKMC), a model that approaches the accuracy of dKMC with a computational cost comparable to conventional hopping. jKMC achieves its computational performance by modelling conduction using identical spherical polarons, yielding a simple delocalisation correction to the Marcus hopping rate that allows polarons to jump over their nearest neighbours. jKMC can be used in regimes of partial delocalisation inaccessible to dKMC to show that modest delocalisation can increase mobilities by as much as two orders of magnitude.
\begin{center}
\includegraphics{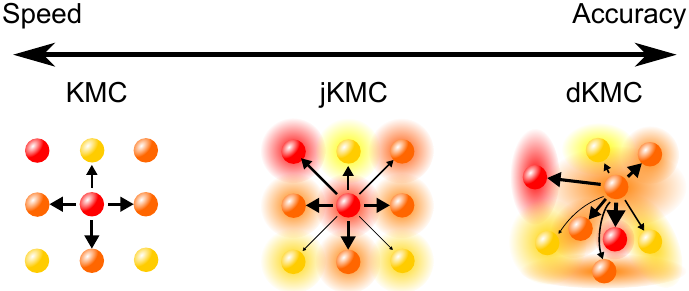}
\end{center}
\end{abstract}

\maketitle

Charge transport is easily modelled in perfectly ordered materials---where charges travel in delocalised bands---and in perfectly disordered ones, where localised charges hop from one site to another. However, many promising materials---such as organic semiconductors, quantum dot arrays, and hybrid perovskites---lie in the intermediate transport regime, where charge transport occurs by hops between partially delocalised states~\cite{Kohler&Bassler,Oberhofer2017,Balzer2021}. In these materials, both static disorder~\cite{Anderson1958} and polaron formation~\cite{Grover1971} localise charges, and partially delocalised states arise when this localisation is insufficient to reduce the state to one molecule. Understanding intermediate-regime transport is especially important in organic semiconductors, where it underpins the most conductive disordered materials and devices~\cite{Giannini2019,Zhang2020}.

State-of-the-art models of partially delocalised charge transport use quantum-mechanical treatments, which can make them computationally expensive. The most detailed models are atomistic simulations such as fragment orbital-based surface hopping~\cite{Spencer2016,Giannini2018,Giannini2019,Giannini2022} and coupled electron-ion dynamics~\cite{Heck2015,Heck2016}. Coarse-graining to reduce computational cost gives effective-Hamiltonian models~\cite{Jiang2016} such as transient localisation~\cite{Fratini2016,Nematiaram2019}, adaptive hierarchy of pure states equations~\cite{Varvelo2021}, density matrix renormalisation group approaches~\cite{Li2020,Li2021}, network approaches~\cite{Savoie2014,Jackson2016}, modified Redfield approaches~\cite{Jankovic2020}, and polaron-transformed Redfield approaches~\cite{Jang2011,Lee2015}. Nevertheless, the computational cost of these models limits them to small systems (usually in one or two dimensions) or short timescales. Recently, we developed delocalised kinetic Monte Carlo (dKMC), a quantum-mechanical model able to describe charge transport in disordered materials on mesoscopic scales and in three dimensions while including the three essential ingredients: disorder, partial delocalisation, and polaron formation~\cite{Balzer2021,Balzer2022}. dKMC demonstrates the importance of delocalisation in disordered charge transport, explaining order-of-magnitude increases in mobility over conventional hopping~\cite{Balzer2021}. However, despite the large computational savings of dKMC compared to other quantum-mechanical treatments, it still remains expensive, making it impractical for simulations of highly delocalised states in three dimensions or on device scales, where the cost scales exponentially with the number of particles.

\begin{figure*}
    \centering
    \includegraphics[width=\textwidth]{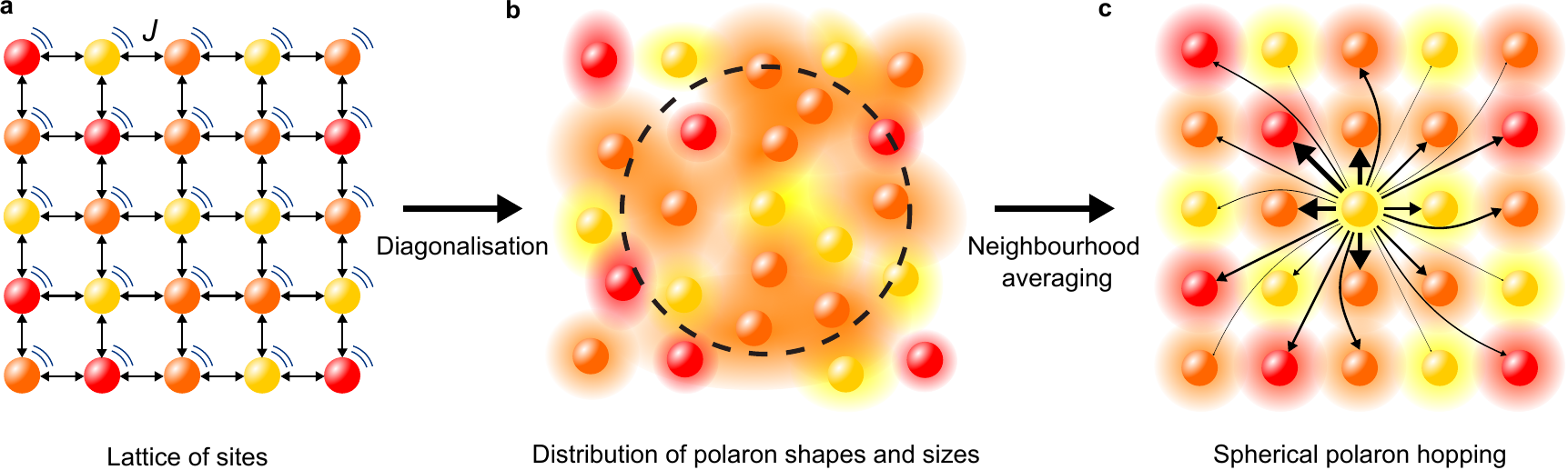}
    \caption{\textbf{The jKMC model of partially delocalised charge transport in disordered materials.} \textbf{(a)} The starting point of jKMC is a lattice of sites with disordered energies (different colours), nearest-neighbour couplings $J$, and coupled to the environment (motion lines). \textbf{(b)} Diagonalising the Hamiltonian yields delocalised polarons with a distribution of shapes and sizes. A suitably chosen neighbourhood (dashed line) is used to average the polaron sizes into a uniform size for jKMC. \textbf{(c)} jKMC uses the neighbourhood-averaged polaron size to represent partially delocalised transport as hopping between uniformly sized spherical polarons. This delocalisation allows polarons to jump over their nearest neighbours.}
    \label{fig:jKMC_model}
\end{figure*}

Here, we develop jumping kinetic Monte Carlo (jKMC), a practical model of intermediate-regime charge transport with speed comparable to hopping models and accuracy comparable to dKMC. We do so by treating the partially delocalised states as a lattice of identical, spherical polarons (\cref{fig:jKMC_model}), allowing us to avoid the most computationally expensive aspects of dKMC. The result of jKMC is a simple delocalisation correction to the Marcus hopping rate that can be included in any transport model. Applied to charge transport, it reveals even greater mobility enhancements for states that are too delocalised to be modelled in dKMC.

Conventional models of charge transport in disordered materials commonly use kinetic Monte Carlo (KMC) simulations to model hopping between an array of sites, usually a lattice. Disorder is commonly introduced by assigning to each site an independent random energy from a Gaussian density of states (DOS) of width $\sigma$~\cite{Bassler1993,Kohler&Bassler}. The probability of each hop and the time taken are determined by the hopping rates, usually expressed using nearest-neighbour Marcus or Miller-Abrahams rates,
\begin{align}
    k_{if}^\mathrm{Marcus}&=\frac{2\pi J^2}{\sqrt{4\pi\lambda k_\mathrm{B}T}}\exp\left(-\frac{\left(E_f-E_i+\lambda\right)^2}{4\lambda k_\mathrm{B}T}\right), 
    \label{eqn:Marcus_rate}\\
    k_{if}^\mathrm{MA}&=
        \begin{cases}
            \nu_0 \,e^{-(E_f-E_i)/k_\mathrm{B}T} & \text{if } E_f>E_i, \\
            \nu_0 & \mathrm{otherwise},
        \end{cases}
    \label{eqn:MA_rate}
\end{align}
where $J$ is the electronic coupling between neighbouring sites, $\lambda$ is the reorganisation energy, $\nu_0$ is the hopping attempt frequency, $T$ is the temperature, $E_i$ and $E_f$ are the energies of the initial and final sites, respectively~\cite{Marcus1956,Miller1960}, and we have set $\hbar=1$. These rates are widely used due to their simplicity, low computational cost, and ability to explain, for example, the electric-field and temperature dependence of mobilities in certain disordered materials~\cite{Bassler1993,Baranovskii2014,Wilken2020,Zojer2021,Upreti2021}.

The nearest-neighbour rates above are often modified to model partially delocalised charge transport by using them to simulate non-nearest-neighbour hopping. To do so, $k_{if}^\mathrm{Marcus}$ and $k_{if}^\mathrm{MA}$ are usually multiplied by a phenomenological tunnelling factor, $e^{-2\gamma d_{if}}$, where $\gamma$ is a fitting parameter called the inverse localisation radius and $d_{if}$ is the hopping distance~\cite{Miller1960,Bassler1993,Zojer2021}. The tunnelling factor was originally developed for impurity conduction in crystalline materials, where it correctly captures the overlap of exponentially decaying tails of distant impurity sites. However, this justification is not valid for densely packed organic semiconductors~\cite{Vukmirovic2010}, where there are many sites in close proximity. However, we will see that jKMC results in an expression similar to the tunnelling factor for realistic amounts of delocalisation in organic semiconductors.

Instead of a phenomenological factor, dKMC models the fundamental processes giving rise to partially delocalised transport~\cite{Balzer2021}. 
dKMC uses the secular polaron-transformed Redfield equation (sPTRE) to model transport between delocalised polarons~\cite{Lee2015}.
It assumes an effective, tight-binding Hamiltonian for a lattice of sites, where each site is linearly coupled to an identical, independent bath of harmonic oscillators. The polaron transformation is applied to this Hamiltonian, reducing the system-bath coupling and allowing it to be treated by second-order perturbative Redfield theory. Applying this treatment to a system with normally distributed energies and nearest-neighbour electronic couplings $J$ yields the hopping rate from polaron $\nu$ into polaron $\nu'$
\begin{multline}
    \label{eqn:sPTRE}
     R_{\nu\nu'}=\sum_{\langle i,j\rangle, \langle i',j'\rangle} 2J^2\mathrm{Re}\big[ \braket{\nu|i}\braket{j|\nu'}\braket{\nu'|i'}\braket{j'|\nu}\\[-0.3cm]\times K_{\Delta(ij,i'j')}(\omega_{\nu\nu'})\big],
\end{multline}
where $\nu$ and $\nu'$ are eigenstates of the polaron-transformed Hamiltonian, $\langle i,j\rangle$ and $\langle i',j'\rangle$ are nearest-neighbour pairs of sites, $\omega_{\nu\nu'}=E_{\nu}-E_{\nu'}$ is the energy difference between the polarons, and $K_{\Delta(ij,i'j')}(\omega)$ describes the residual system-bath coupling in the polaron frame, as described in \cref{app:jKMC_rate_derivation}. In calculating $K_{\Delta(ij,i'j')}(\omega)$, we assume a super-Ohmic spectral density $J(\omega)=\frac{\pi\lambda}{4}(\omega/\omega_c)^3e^{-\omega/\omega_c}$, where $\lambda$ is the reorganisation energy and the cutoff frequency is set to $\omega_c=\SI{62}{meV}$~\cite{Lee2015}.

Calculating the hopping rate $R_{\nu\nu'}$ in dKMC requires diagonalising disordered Hamiltonians to calculate the delocalised polaron states $\ket{\nu}$, which have irregular shapes and off-lattice positions. Diagonalising these Hamiltonians becomes the computational bottleneck for dKMC in three dimensions, and at large $J$ the states become too large to be contained within a Hamiltonian that can be diagonalised.

jKMC avoids the computational bottleneck of dKMC by avoiding the calculation of all the polaron states. Instead, it assumes that the polaron wavefunctions are identical and spherically symmetric. We also assume that the polarons are centred on a cubic lattice with spacing $a$, have independent and normally distributed energies (with mean 0 and standard deviation $\sigma$), and that their shapes follow the exponential localisation seen in the Anderson model~\cite{Anderson1958}. Specifically, we take
\begin{equation}
    \label{eqn:Spherical_polaron_approximation}
    \ket{\nu}=A \sum_i  \exp\left(-\frac{d_{i \nu}}{r_\mathrm{deloc}}\right)\ket{i},
\end{equation}
where $d_{i \nu}$ is the distance between the centre of the polaron $\nu$ and site $i$, $r_\mathrm{deloc}$ is the delocalisation radius that characterises the size of the wavefunction, and $A=\left(\sum_i \exp\left(-2d_{i \nu}/r_\mathrm{deloc}\right)\right)^{-1/2}$ is the normalisation. 

\begin{figure}
    \centering
    \includegraphics[width=\columnwidth]{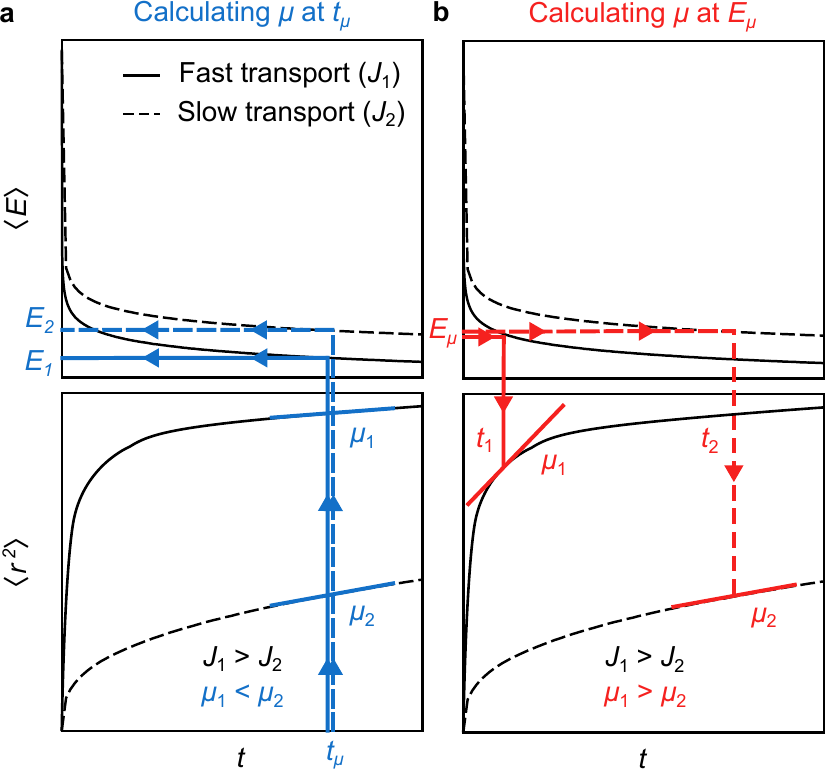}
    \caption{\textbf{Calculating mobilities at a target energy.} 
    \textbf{(a)}~Conventionally, mobilities in dispersive materials are calculated at a particular time $t_\mu$ from the slope of the mean-squared displacement $\langle r^2 \rangle$ as a function of time. This approach can unrealistically predict smaller mobilities for carriers with higher electronic couplings $J$ because polarons with larger $J$ can move faster and reach deeper traps in the DOS ($E_1 < E_2$) within a given $t_\mu$.
    \textbf{(b)}~This problem is avoided by calculating mobilities at a particular energy $E_\mu$. Here, the mobilities are calculated at the times $t_1$ and $t_2$ when the respective simulations reach the target energy.} 
    \label{fig:Mobility_comparison}
\end{figure}

Parametrising jKMC requires choosing a value of $r_\mathrm{deloc}$. Our objective is to choose the $r_\mathrm{deloc}$ which will yield accurate mobilities. In disordered materials, $r_\mathrm{deloc}$ should depend on the mean energy $\langle E \rangle$ of the polarons because polaron sizes decrease (on average) as they relax to more localised states lower in the disordered DOS.
For example, $r_\mathrm{deloc}$ should be larger for an ensemble of randomly occupied polaron states (where $\langle E \rangle=0$ and many large states in the middle of the DOS are occupied) than for an ensemble of polarons that have reached thermal equilibrium (where $\langle E \rangle=-\sigma^2/k_\mathrm{B}T$ and the occupied states are mostly the localised traps)~\cite{Kohler&Bassler}.

To choose $r_\mathrm{deloc}$, we relate it to a readily calculated measure of delocalisation, the inverse participation ratio
\begin{equation}
    \label{eqn:IPR}
    \mathrm{IPR}_\nu=\Big(\displaystyle \sum_i \abs{\braket{i|\nu}}^4\Big)^{-1},
\end{equation}
which roughly equals the number of sites over which polaron $\nu$ is delocalised. A localised wavefunction has $\mathrm{IPR}=1$, while a wavefunction evenly spread across $N$ sites ($\braket{i|\nu}=N^{-1/2}$) has $\mathrm{IPR}=N$. The spherical polarons of \cref{eqn:Spherical_polaron_approximation} have an IPR of
\begin{equation}
    \label{eqn:IPR-rdeloc_conversion}
    \mathrm{IPR}_\mathrm{jKMC}=A^{-4}\Bigg(\sum_i \exp \left(-\frac{4d_{i\nu}}{r_\mathrm{deloc}}\Bigg)\right)^{-1},
\end{equation}
an equation that allows us to calculate an $r_\mathrm{deloc}$ that reproduces a given IPR.

We set $r_\mathrm{deloc}$ using \cref{eqn:IPR-rdeloc_conversion} based on the mean IPR of the polaron states that participate in charge transport at a given $\langle E \rangle$. This IPR is calculated through an approach we call neighbourhood averaging.
First, we note that the averaging of the polaron IPRs should be thermally weighted because transport is driven by relaxation to thermal equilibrium.
A complete thermal average requires that every state be accessible; however, during the initial stages of transport in a disordered material, the polaron is unable to completely explore the DOS. Instead, if a polaron can only explore a local neighbourhood of $N$ polaron states until a particular time, we take the thermal averages of polaron states within that neighbourhood. Therefore, we define the effective IPR and the effective energy as
\begin{align}
    \mathrm{IPR}_\mathrm{eff}(N)&=\left\langle\frac{1}{Z} \sum_{\nu=1}^N \mathrm{IPR}_\nu \exp\left(-\frac{E_\nu}{k_\mathrm{B}T}\right)\right\rangle,
    \label{eqn:Effective_IPR}\\
    E_\mathrm{eff}(N)&=\left\langle\frac{1}{Z} \sum_{\nu=1}^N E_\nu \exp\left(-\frac{E_\nu}{k_\mathrm{B}T}\right)\right\rangle,
    \label{eqn:Effective_E}
\end{align}
where $\mathrm{IPR}_\nu$ and $E_\nu$ are the polaron IPRs and energies obtained from the diagonalisation of the model Hamiltonian~\cite{Balzer2021}, $Z=\sum_{\nu=1}^N \exp\left(-E_\nu/k_\mathrm{B}T\right)$ is the partition function, and the average $\langle \cdot \rangle$ is taken over an ensemble of disordered energetic landscapes (1000 instances in our calculations). 
\Cref{eqn:Effective_IPR,eqn:Effective_E} allow us to obtain an $r_\mathrm{deloc}$ for any $\langle E \rangle$ in two steps. First, for a given $\langle E \rangle$, we determine the appropriate neighbourhood size $N$ using \cref{eqn:Effective_E} and second, we use that same $N$ in \cref{eqn:Effective_IPR} to determine the $\mathrm{IPR}_\mathrm{eff}$ that can be converted into $r_\mathrm{deloc}$ using \cref{eqn:IPR-rdeloc_conversion}.

With $r_\mathrm{deloc}$ in hand, we can now substitute the spherical-polaron approximation in \cref{eqn:Spherical_polaron_approximation} into the delocalised polaron hopping rate, \cref{eqn:sPTRE}.
In order to obtain a simple rate expression, we also assume the high-temperature limit ($k_\mathrm{B}T\gg \omega_c$), because many organic semiconductors operate close to this limit~\cite{Bredas2002,Hoffmann2012,Shuai2020}. In addition, the high-temperature limit is the regime of validity of ordinary Marcus theory and using it reveals the relationship between ordinary KMC and jKMC. Using the spherical-polaron and high-temperature approximations, we obtain the jKMC rate between any two polarons (derivation in \cref{app:jKMC_rate_derivation}),
\begin{equation}
    \label{eqn:jKMC_rate_Marcus}
    k_{\nu\nu'}^{\mathrm{jKMC}}=k_{\nu\nu'}^\mathrm{Marcus}\xi_{\nu\nu'},
\end{equation}
where $k_{\nu\nu'}^\mathrm{Marcus}$ is the Marcus rate of \cref{eqn:Marcus_rate} from polaron $\nu$ to $\nu'$ as if they were nearest neighbours and $\xi_{\nu\nu'}$ is the delocalisation correction
\begin{equation}
    \label{eqn:Delocalisation_correction}
    \xi_{\nu\nu'}=  A^4\sum_{\langle i,j\rangle} \exp\left(-\frac{2(d_{i\nu }+d_{j\nu'})}{r_\mathrm{deloc}}\right),
\end{equation}
where the sum runs over nearest-neighbour pairs of sites $i$ and $j$. Hence, the effect of delocalisation is to make Marcus rates long range in a way that depends straightforwardly on the delocalisation radius $r_\mathrm{deloc}$.

In the low-delocalisation limit, $\xi_{\nu\nu'}$ can be simplified by taking only the dominant exponential terms in \cref{eqn:Delocalisation_correction}, which leads to the simplified jKMC rate (derivation in \cref{app:Simplified_jKMC_rate_derivation})
\begin{equation}
    \label{eqn:Simplified_jKMC_rate}
    \xi_{\nu\nu'}^\mathrm{Simplified}=\frac{d_{\nu\nu'}}{a}\exp\left(-\frac{2\left(d_{\nu\nu'}-a\right)}{r_\mathrm{deloc}}\right).
\end{equation}
Simplified jKMC is similar to inserting the phenomenological correction $e^{-2\gamma d_{if}}$ into the Marcus rate, and can be thought of as a rigorous justification of this correction in the limit of small delocalisation. However, there are two important differences. First, the distance-dependent exponent is offset by the lattice constant, which ensures that the correct Marcus rate is reproduced for nearest neighbours. Second, there is a new pre-exponential factor $d_{\nu\nu'}/a$ which accounts for the number of terms significantly contributing to the sum in \cref{eqn:Delocalisation_correction}. 

The jKMC rates $k_{\nu\nu'}^{\mathrm{jKMC}}$ can now be used to simulate polaron dynamics. Disordered materials show dispersive transport (mobilities decreasing over time as the polarons relax within the DOS)~\cite{Hoffmann2012,Melianas2014,Melianas2019,Upreti2021}, which affects how mobilities should be calculated. Conventionally, mobilities in dispersive systems have been calculated at a chosen time $t_\mu$ (\cref{fig:Mobility_comparison}a); however, this approach can lead to unrealistic comparisons, because it can predict lower mobilities for systems with stronger couplings $J$, where polarons can reach deeper traps in the DOS within the same $t_\mu$. To avoid this pitfall, we calculate mobilities of polarons that have relaxed (on average) to a chosen target energy $E_\mu$ (i.e., for which $\langle E \rangle = E_\mu$, see \cref{fig:Mobility_comparison}b). In particular, in jKMC, we choose the value of $r_\mathrm{deloc}$ that is consistent with this target $E_\mu$.

\begin{figure}
    \centering
    \includegraphics[width=\columnwidth]{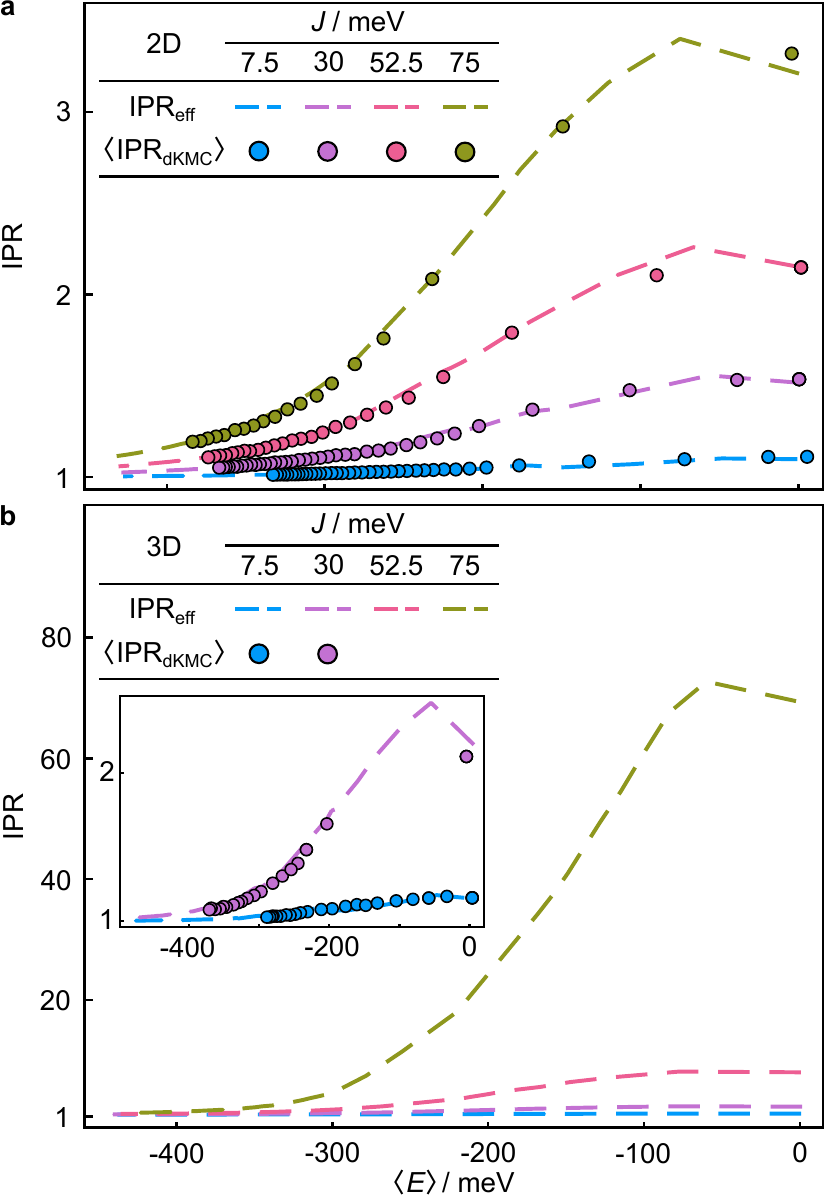}
    \caption{\textbf{The neighbourhood-averaging approach predicts accurate IPRs.} 
    In both \textbf{(a)} two and \textbf{(b)} three dimensions, the neighbourhood-averaging approach gives the effective IPR (dashed lines) as a function of the mean polaron energy $\langle E\rangle$.  
    These results reproduce the mean IPR of occupied states in dKMC (points) obtained using fully dynamical simulations of polarons relaxing in the DOS. The right-most point corresponds to time $t=0$, and subsequent points (towards lower $\langle E \rangle$ and lower IPR) correspond to progressively longer dKMC simulations. Results are calculated for $\sigma=\SI{150}{meV}$, $\lambda=\SI{200}{meV}$, and $T=\SI{300}{K}$.}
    \label{fig:IPR_E}
\end{figure}

\begin{figure*}
    \centering
    \includegraphics[width=\textwidth]{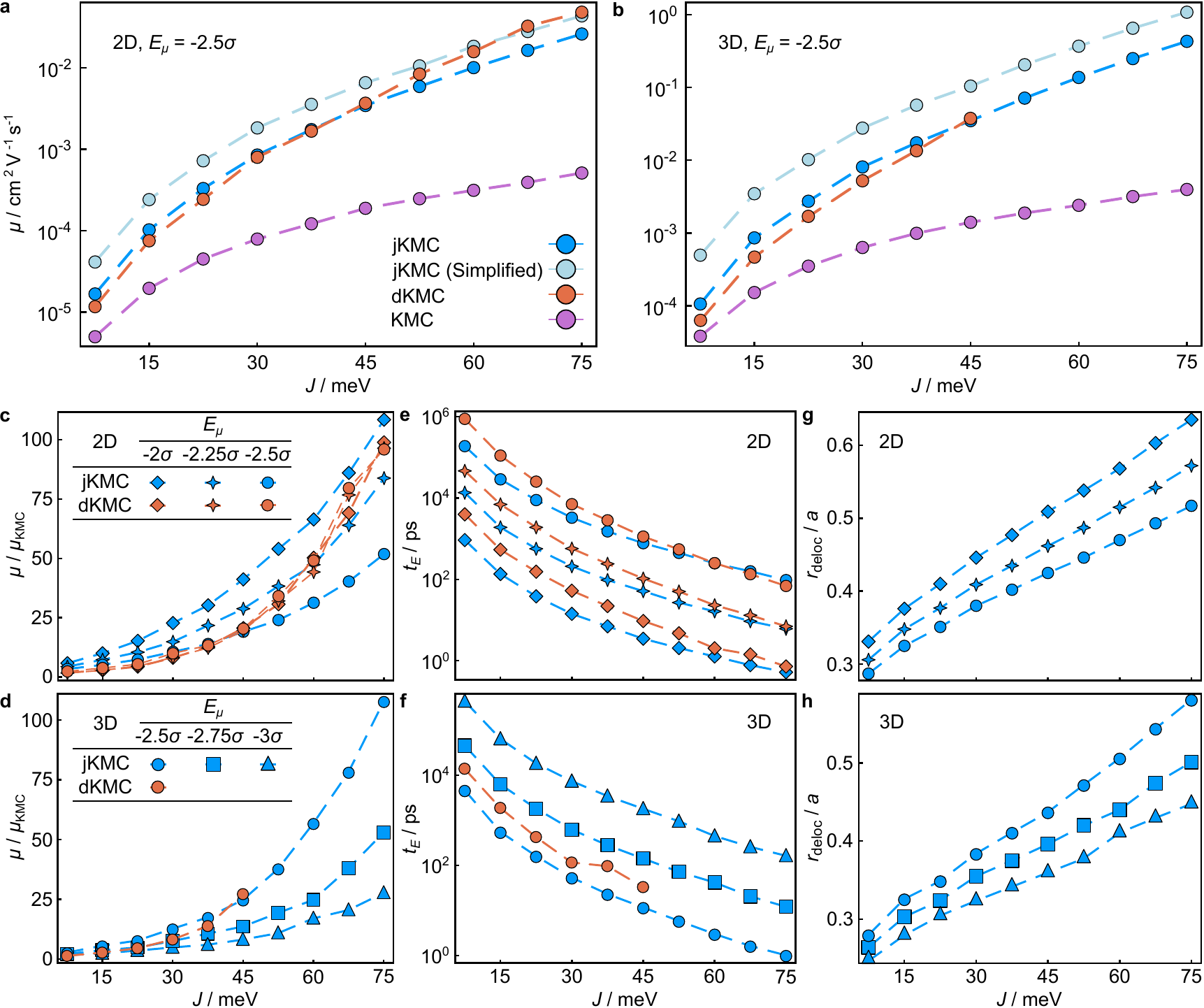}
    \caption{\textbf{jKMC reproduces the large delocalisation enhancements explained by dKMC},
    including the two-order-of-magnitude enhancements in mobility over KMC in \textbf{(a)} two and \textbf{(b)} three dimensions. jKMC reproduces dKMC mobilities where the latter are known, and extends beyond them to high delocalisation in three dimensions. The simplified jKMC rate also predicts mobilities on the same order of magnitude as dKMC. Mobilities are calculated at $E_\mu=-2.5\sigma$ with $\sigma=\SI{150}{meV}$, $\lambda=\SI{200}{meV}$, and $T=\SI{300}{K}$. \textbf{(c,d)} jKMC reproduces the delocalisation enhancements of dKMC over KMC across a range of $E_\mu$ to about a factor of 2. \textbf{(e,f)} The times $t_E$ taken to reach the target energy $E_\mu$ show that the range of $E_\mu$ in (c,d) spans the typical timescales of mesoscopic charge transport. \textbf{(g,h)} The delocalisation radii that parametrise jKMC.}
    \label{fig:Results}
\end{figure*}

To simulate transport, we initialise a polaron in the centre of the lattice, and calculate 10 KMC trajectories~\cite{Bassler1993,Kohler&Bassler} over \num{10000} disordered landscapes using the jKMC rate. During these simulations, we track the squared displacement $r^2(t)$ of the polaron and its energy $E(t)$.
We average these quantities over the ensemble of trajectories to obtain the mean squared displacement $\langle r^2 (t)\rangle$ and the mean energy $\langle E(t)\rangle$ of the diffusing polaron. We then calculate the mobility at the time $t_E$ at which $\langle E(t_E) \rangle=E_\mu$ using
\begin{equation}
    \mu(t_E)=\left.\frac{e}{2d\,k_\mathrm{B}T}  \frac{d}{dt}\langle r^2(t) \rangle\right|_{t=t_E},
    \label{eqn:Mobility}
\end{equation}
where $e$ is the electron charge and $d$ is the dimension.

Before discussing jKMC mobilities, we show that the neighbourhood-averaging approach reproduces the mean IPRs obtained from full dKMC calculations (\cref{fig:IPR_E}).
In dKMC, the mean IPR $\langle \mathrm{IPR}_\mathrm{dKMC} \rangle$ and mean energy $\langle E_\mathrm{dKMC} \rangle$ of the occupied states can be obtained as functions of time by simulating and averaging transport over ensembles of diagonalised polaron landscapes. 
Therefore, \cref{fig:IPR_E} shows that the neighbourhood-averaging approach can be used to predict the mean IPR (and, therefore, $r_\mathrm{deloc}$) across a wide range of parameters without expensive dKMC calculations.

\begin{figure}
    \centering
    \includegraphics[width=\columnwidth]{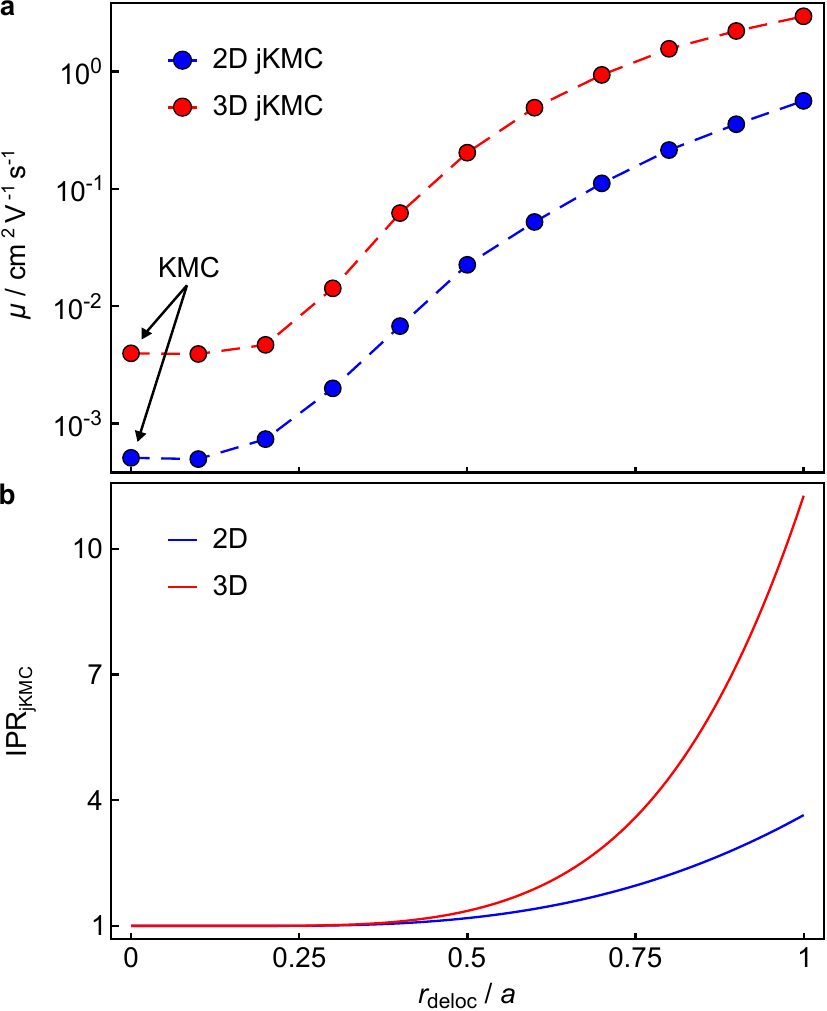}
    \caption{\textbf{The jKMC delocalisation correction can produce large increases in mobilities.}
    \textbf{(a)} In both two and three dimensions, varying $r_\mathrm{deloc}$ independently of $J$ increases the mobility by several orders of magnitude. Therefore, adding the delocalisation correction $\xi_{\nu\nu'}$ can easily reproduce large mobilities where conventional KMC hopping fails.
    Results are calculated at $E_\mu=-2.5\sigma$ for $J=\SI{75}{meV}$, $\sigma=\SI{150}{meV}$, $\lambda=\SI{200}{meV}$, and $T=\SI{300}{K}$. \textbf{(b)} The jKMC IPR only depends on $r_\mathrm{deloc}$, and can be varied to probe the effect of delocalisation. Modest amounts of delocalisation produce order-of-magnitude enhancements in panel \textbf{a}.}
    \label{fig:Results_r}
\end{figure}

jKMC reproduces the large delocalisation enhancements seen in dKMC over a wide range of electronic couplings (\cref{fig:Results}a,b). In particular, for the parameters chosen, it reproduces the two-order-of-magnitude enhancement over KMC. jKMC can also describe regimes inaccessible to dKMC, especially transport involving large electronic couplings in three dimensions, where it predicts even larger enhancements in mobility.

The delocalisation enhancements of jKMC remain large over a range of $E_\mu$ in both two and three dimensions (\Cref{fig:Results}c,d). The chosen range of $E_\mu$ corresponds to several-orders-of-magnitude differences in the transport time $t_E$ (\cref{fig:Results}e,f), showing that the delocalisation enhancement persists over a broad range of timescales. For the parameters tested, jKMC agrees with dKMC to about a factor of 2. Overall, jKMC provides excellent accuracy considering the simplicity of the method, typical uncertainties in the input parameters, and the fact that the mobilities span orders of magnitude and are underestimated by KMC by as much as 100 times.

The values of $r_\mathrm{deloc}$ used in jKMC (\cref{fig:Results}g,h) are within the typical range of its analogue $\gamma^{-1}$ sometimes used in Marcus or Miller-Abrahams models. For the parameters used in \cref{fig:Results}, we predict $r_\mathrm{deloc}/a$ within a range of 0.2--0.6, whereas the values of $\gamma^{-1}/a$ are typically 0.1--1~\cite{Hoffmann2012,Baranovskii2014}. In other words, jKMC is a microscopic justification of phenomenological $\gamma^{-1}$ parameters that are often inserted into hopping models to enable them to fit experimental results.

\Cref{fig:Results_r}a shows that the delocalisation correction can increase mobilities by several orders of magnitude over KMC hopping ($r_\mathrm{deloc}=0$). In \cref{fig:Results_r}a, $r_\mathrm{deloc}$ is varied independently, isolating its role from the other material parameters. The IPRs corresponding to $r_\mathrm{deloc}$ are shown in \cref{fig:Results_r}b, which remain modest for the significant enhancements in mobility.

The results of simplified jKMC are also shown in \cref{fig:Results}a,b (and \cref{app:Simplified_jKMC_results}), showing that \cref{eqn:Simplified_jKMC_rate} can be an acceptable approximation for typical delocalisations in disordered organic semiconductors. The simplified jKMC rate leads to the same order-of-magnitude mobilities as dKMC; however, it tends to slightly systematically overestimate jKMC. 

Despite its overall accuracy and the ability to predict the correct trends, jKMC has systematic errors in certain regimes. 
\Cref{fig:Results}c,d shows jKMC overestimates mobilities on short timescales (shallow $E_\mu$) and underestimates them at long timescales (deep $E_\mu$). These errors are related to the assumption of uniform polaron size, which neglects the effect of the distribution of polaron sizes on the mobility. At short timescales, where the distribution is wide and the effective IPR is large, jKMC overestimates the escape from localised traps. In contrast, on long timescales where the distribution has narrowed and the effective IPR is small, jKMC underestimates the escape from traps into highly delocalised states. The boundary between these regimes depends on the coupling $J$, as shown in \cref{fig:Results}c,d.

jKMC achieves a computational cost between those of KMC and dKMC. Although based on dKMC, jKMC avoids dKMC's computational bottleneck of having to repeatedly diagonalise Hamiltonians during the dynamics. Instead, it requires only an up-front diagonalisation to calculate the effective IPR. Furthermore, jKMC can achieve the neighbourhood averaging by diagonalising fewer and smaller lattices than need to be diagonalised in dKMC. Smaller lattices suffice in jKMC because the effective IPR is heavily weighted towards smaller, lower-energy polarons (see \cref{app:Effective_IPR_convergence}). In contrast, dKMC is designed to calculate the dynamics of any polaron on the lattice, including highly delocalised states which require large lattices to describe. Nevertheless, the up-front diagonalisation remains the computational bottleneck of jKMC at high electronic couplings. However, it only needs to be done once before running a routine KMC calculation using jKMC rates.

The approximations in jKMC allow it to demonstrate the important role of delocalisation in a wide range of organic semiconductors, and the simplicity of the resulting equations gives clear insight into how the microscopic parameters affect the mesoscopic dynamics. However, most of the approximations could be relaxed in order to extend jKMC to other regimes.

The distinguishing assumption of jKMC is the uniformly sized spherical polarons, and both the assumptions of uniform size and shape could be relaxed.
Polaron sizes could be made non-uniform to capture the distribution of polaron IPRs and their correlation to the polaron energies (i.e., lower-energy states are generally more localised than those in the middle of the DOS). This could be achieved by using a different $r_\mathrm{deloc}$ for each polaron, as a function of its energy. However, doing so would require both keeping track of the polaron IPR distribution and modifications to hopping rates to account for different initial and final polaron $r_\mathrm{deloc}$s.
The assumption of spherical size could likewise be relaxed without using the raw polaron wavefunctions obtained by diagonalisation. Spherical polarons are appropriate for materials with isotropic couplings; by contrast, ellipsoidal polarons could be implemented to describe anisotropic couplings, such as those in polymers or oligoacene crystals. 

The high-temperature assumption could also be relaxed, at the cost of losing the simplicity and intuition of jKMC. In particular, it is possible to directly use the sPTRE rates in \cref{eqn:sPTRE} with the spherical polarons of jKMC but without taking the high-temperature limit of $K_{\Delta(ij,i'j')}(\omega)$. These equations are given as intermediate results in \cref{app:jKMC_rate_derivation}.

Other approximations in jKMC are shared by the implementations of the underlying sPTRE and dKMC theories. If they were relaxed, the sPTRE equations of motion might change, which would imply consequent changes in jKMC dynamics. For example, it would be possible to extend jKMC to accommodate long-range electronic couplings, a non-Gaussian DOS~\cite{Vissenberg1998,Vukmirovic2010,Baranovskii2014}, or spatial site-energy correlations~\cite{Gartstein1995,Baranovskii2014}, regimes that are rarely explored with sPTRE or dKMC. There are also approximations that are inherited from sPTRE and dKMC. For example, sPTRE assumes fixed electronic couplings, which could be relaxed to include fluctuations that can dominate transport in energetically ordered materials~\cite{Troisi2006,Troisi2006_2}.  

In the future, jKMC opens the opportunity for determining the effect of delocalisation on other optoelectronic processes in disordered materials or on a device level, regimes that would be too complicated to explore using dKMC or any other quantum-mechanical method~\cite{Balzer2022}. For example, we expect that jKMC can be extended to modelling multiple particles, including charge separation and recombination processes. These simulations could be used to parametrise drift-diffusion simulations of delocalised charges, connecting the mesoscopic dynamics to a complete, multi-scale device model.

In conclusion, jKMC is a practical model of partially delocalised transport that approaches the accuracy of fully quantum approaches with the cost of classical hopping. jKMC includes a simple correction to the Marcus hopping rate, based on a method to estimate the delocalisation radius. It reproduces the large increases in mobility predicted by dKMC, but can also simulate larger electronic couplings and delocalisation in three dimensions. These factors make jKMC an attractive model that could easily be included in any KMC simulation of a disordered material, including future device-scale models that take into account partially delocalised charge transport.

\begin{acknowledgments}
We were supported by the Australian Research Council (DP220103584), by a Westpac Scholars Trust Future Leaders Scholarship, by an Australian Government Research Training Program scholarship, and by computational resources from the Sydney Informatics Hub (Artemis) and the Australian Government's National Computational Infrastructure (Gadi) through the National Computational Merit Allocation Scheme.
\end{acknowledgments}

\bibliography{bib}

\begin{thebibliography}{41}%
\makeatletter
\providecommand \@ifxundefined [1]{%
 \@ifx{#1\undefined}
}%
\providecommand \@ifnum [1]{%
 \ifnum #1\expandafter \@firstoftwo
 \else \expandafter \@secondoftwo
 \fi
}%
\providecommand \@ifx [1]{%
 \ifx #1\expandafter \@firstoftwo
 \else \expandafter \@secondoftwo
 \fi
}%
\providecommand \natexlab [1]{#1}%
\providecommand \enquote  [1]{``#1''}%
\providecommand \bibnamefont  [1]{#1}%
\providecommand \bibfnamefont [1]{#1}%
\providecommand \citenamefont [1]{#1}%
\providecommand \href@noop [0]{\@secondoftwo}%
\providecommand \href [0]{\begingroup \@sanitize@url \@href}%
\providecommand \@href[1]{\@@startlink{#1}\@@href}%
\providecommand \@@href[1]{\endgroup#1\@@endlink}%
\providecommand \@sanitize@url [0]{\catcode `\\12\catcode `\$12\catcode
  `\&12\catcode `\#12\catcode `\^12\catcode `\_12\catcode `\%12\relax}%
\providecommand \@@startlink[1]{}%
\providecommand \@@endlink[0]{}%
\providecommand \url  [0]{\begingroup\@sanitize@url \@url }%
\providecommand \@url [1]{\endgroup\@href {#1}{\urlprefix }}%
\providecommand \urlprefix  [0]{URL }%
\providecommand \Eprint [0]{\href }%
\providecommand \doibase [0]{https://doi.org/}%
\providecommand \selectlanguage [0]{\@gobble}%
\providecommand \bibinfo  [0]{\@secondoftwo}%
\providecommand \bibfield  [0]{\@secondoftwo}%
\providecommand \translation [1]{[#1]}%
\providecommand \BibitemOpen [0]{}%
\providecommand \bibitemStop [0]{}%
\providecommand \bibitemNoStop [0]{.\EOS\space}%
\providecommand \EOS [0]{\spacefactor3000\relax}%
\providecommand \BibitemShut  [1]{\csname bibitem#1\endcsname}%
\let\auto@bib@innerbib\@empty
\bibitem [{\citenamefont {K\"ohler}\ and\ \citenamefont
  {B\"assler}(2015)}]{Kohler&Bassler}%
  \BibitemOpen
  \bibfield  {author} {\bibinfo {author} {\bibfnamefont {A.}~\bibnamefont
  {K\"ohler}}\ and\ \bibinfo {author} {\bibfnamefont {H.}~\bibnamefont
  {B\"assler}},\ }\href@noop {} {\emph {\bibinfo {title} {Electronic Processes
  in Organic Semiconductors}}},\ \bibinfo {edition} {1st}\ ed.\ (\bibinfo
  {publisher} {Wiley-VCH},\ \bibinfo {year} {2015})\BibitemShut {NoStop}%
\bibitem [{\citenamefont {Oberhofer}\ \emph {et~al.}(2017)\citenamefont
  {Oberhofer}, \citenamefont {Reuter},\ and\ \citenamefont
  {Blumberger}}]{Oberhofer2017}%
  \BibitemOpen
  \bibfield  {author} {\bibinfo {author} {\bibfnamefont {H.}~\bibnamefont
  {Oberhofer}}, \bibinfo {author} {\bibfnamefont {K.}~\bibnamefont {Reuter}},\
  and\ \bibinfo {author} {\bibfnamefont {J.}~\bibnamefont {Blumberger}},\
  }\bibfield  {title} {\bibinfo {title} {Charge transport in molecular
  materials: {A}n assessment of computational methods},\ }\href
  {https://doi.org/10.1021/acs.chemrev.7b00086} {\bibfield  {journal} {\bibinfo
   {journal} {Chem. Rev.}\ }\textbf {\bibinfo {volume} {117}},\ \bibinfo
  {pages} {10319} (\bibinfo {year} {2017})}\BibitemShut {NoStop}%
\bibitem [{\citenamefont {Balzer}\ \emph {et~al.}(2021)\citenamefont {Balzer},
  \citenamefont {Smolders}, \citenamefont {Blyth}, \citenamefont {Hood},\ and\
  \citenamefont {Kassal}}]{Balzer2021}%
  \BibitemOpen
  \bibfield  {author} {\bibinfo {author} {\bibfnamefont {D.}~\bibnamefont
  {Balzer}}, \bibinfo {author} {\bibfnamefont {T.~J. A.~M.}\ \bibnamefont
  {Smolders}}, \bibinfo {author} {\bibfnamefont {D.}~\bibnamefont {Blyth}},
  \bibinfo {author} {\bibfnamefont {S.~N.}\ \bibnamefont {Hood}},\ and\
  \bibinfo {author} {\bibfnamefont {I.}~\bibnamefont {Kassal}},\ }\bibfield
  {title} {\bibinfo {title} {Delocalised kinetic {M}onte {C}arlo for simulating
  delocalisation-enhanced charge and exciton transport in disordered
  materials},\ }\href {https://doi.org/10.1039/Balzer2020} {\bibfield
  {journal} {\bibinfo  {journal} {Chem. Sci.}\ }\textbf {\bibinfo {volume}
  {12}},\ \bibinfo {pages} {2276} (\bibinfo {year} {2021})}\BibitemShut
  {NoStop}%
\bibitem [{\citenamefont {Anderson}(1958)}]{Anderson1958}%
  \BibitemOpen
  \bibfield  {author} {\bibinfo {author} {\bibfnamefont {P.}~\bibnamefont
  {Anderson}},\ }\bibfield  {title} {\bibinfo {title} {Absence of diffusion in
  certain random lattices},\ }\href {https://doi.org/10.1103/PhysRev.109.1492}
  {\bibfield  {journal} {\bibinfo  {journal} {Phys. Rev.}\ }\textbf {\bibinfo
  {volume} {109}},\ \bibinfo {pages} {1492} (\bibinfo {year}
  {1958})}\BibitemShut {NoStop}%
\bibitem [{\citenamefont {Grover}\ and\ \citenamefont
  {Silbey}(1971)}]{Grover1971}%
  \BibitemOpen
  \bibfield  {author} {\bibinfo {author} {\bibfnamefont {M.}~\bibnamefont
  {Grover}}\ and\ \bibinfo {author} {\bibfnamefont {R.}~\bibnamefont
  {Silbey}},\ }\bibfield  {title} {\bibinfo {title} {Exciton migration in
  molecular crystals},\ }\href {https://doi.org/10.1063/1.1674761} {\bibfield
  {journal} {\bibinfo  {journal} {J. Chem. Phys.}\ }\textbf {\bibinfo {volume}
  {54}},\ \bibinfo {pages} {4843} (\bibinfo {year} {1971})}\BibitemShut
  {NoStop}%
\bibitem [{\citenamefont {Giannini}\ \emph {et~al.}(2019)\citenamefont
  {Giannini}, \citenamefont {Carof}, \citenamefont {Ellis}, \citenamefont
  {Yang}, \citenamefont {Ziogos}, \citenamefont {Ghosh},\ and\ \citenamefont
  {Blumberger}}]{Giannini2019}%
  \BibitemOpen
  \bibfield  {author} {\bibinfo {author} {\bibfnamefont {S.}~\bibnamefont
  {Giannini}}, \bibinfo {author} {\bibfnamefont {A.}~\bibnamefont {Carof}},
  \bibinfo {author} {\bibfnamefont {M.}~\bibnamefont {Ellis}}, \bibinfo
  {author} {\bibfnamefont {H.}~\bibnamefont {Yang}}, \bibinfo {author}
  {\bibfnamefont {O.~G.}\ \bibnamefont {Ziogos}}, \bibinfo {author}
  {\bibfnamefont {S.}~\bibnamefont {Ghosh}},\ and\ \bibinfo {author}
  {\bibfnamefont {J.}~\bibnamefont {Blumberger}},\ }\bibfield  {title}
  {\bibinfo {title} {Quantum localization and delocalization of charge carriers
  in organic semiconducting crystals},\ }\href
  {https://doi.org/10.1038/s41467-019-11775-9} {\bibfield  {journal} {\bibinfo
  {journal} {Nat. Commun.}\ }\textbf {\bibinfo {volume} {10}},\ \bibinfo
  {pages} {3843} (\bibinfo {year} {2019})}\BibitemShut {NoStop}%
\bibitem [{\citenamefont {Zhang}\ \emph {et~al.}(2020)\citenamefont {Zhang},
  \citenamefont {Chen}, \citenamefont {Xiao}, \citenamefont {Chow},
  \citenamefont {Ren}, \citenamefont {Kupgan}, \citenamefont {Jiao},
  \citenamefont {Chan}, \citenamefont {Du}, \citenamefont {Xia}, \citenamefont
  {Chen}, \citenamefont {Yuan}, \citenamefont {Zhang}, \citenamefont {Zhang},
  \citenamefont {Liu}, \citenamefont {Zou}, \citenamefont {Yan}, \citenamefont
  {Wong}, \citenamefont {Coropceanu}, \citenamefont {Li}, \citenamefont
  {Brabec}, \citenamefont {Bredas}, \citenamefont {Yip},\ and\ \citenamefont
  {Cao}}]{Zhang2020}%
  \BibitemOpen
  \bibfield  {author} {\bibinfo {author} {\bibfnamefont {G.}~\bibnamefont
  {Zhang}}, \bibinfo {author} {\bibfnamefont {X.-K.}\ \bibnamefont {Chen}},
  \bibinfo {author} {\bibfnamefont {J.}~\bibnamefont {Xiao}}, \bibinfo {author}
  {\bibfnamefont {P.~C.~Y.}\ \bibnamefont {Chow}}, \bibinfo {author}
  {\bibfnamefont {M.}~\bibnamefont {Ren}}, \bibinfo {author} {\bibfnamefont
  {G.}~\bibnamefont {Kupgan}}, \bibinfo {author} {\bibfnamefont
  {X.}~\bibnamefont {Jiao}}, \bibinfo {author} {\bibfnamefont {C.~C.~S.}\
  \bibnamefont {Chan}}, \bibinfo {author} {\bibfnamefont {X.}~\bibnamefont
  {Du}}, \bibinfo {author} {\bibfnamefont {R.}~\bibnamefont {Xia}}, \bibinfo
  {author} {\bibfnamefont {Z.}~\bibnamefont {Chen}}, \bibinfo {author}
  {\bibfnamefont {J.}~\bibnamefont {Yuan}}, \bibinfo {author} {\bibfnamefont
  {Y.}~\bibnamefont {Zhang}}, \bibinfo {author} {\bibfnamefont
  {S.}~\bibnamefont {Zhang}}, \bibinfo {author} {\bibfnamefont
  {Y.}~\bibnamefont {Liu}}, \bibinfo {author} {\bibfnamefont {Y.}~\bibnamefont
  {Zou}}, \bibinfo {author} {\bibfnamefont {H.}~\bibnamefont {Yan}}, \bibinfo
  {author} {\bibfnamefont {K.~S.}\ \bibnamefont {Wong}}, \bibinfo {author}
  {\bibfnamefont {V.}~\bibnamefont {Coropceanu}}, \bibinfo {author}
  {\bibfnamefont {N.}~\bibnamefont {Li}}, \bibinfo {author} {\bibfnamefont
  {C.~J.}\ \bibnamefont {Brabec}}, \bibinfo {author} {\bibfnamefont {J.-L.}\
  \bibnamefont {Bredas}}, \bibinfo {author} {\bibfnamefont {H.-L.}\
  \bibnamefont {Yip}},\ and\ \bibinfo {author} {\bibfnamefont {Y.}~\bibnamefont
  {Cao}},\ }\bibfield  {title} {\bibinfo {title} {Delocalization of exciton and
  electron wavefunction in non-fullerene acceptor molecules enables efficient
  organic solar cells},\ }\href {https://doi.org/10.1038/s41467-020-17867-1}
  {\bibfield  {journal} {\bibinfo  {journal} {Nat. Commun.}\ }\textbf {\bibinfo
  {volume} {11}},\ \bibinfo {pages} {3943} (\bibinfo {year}
  {2020})}\BibitemShut {NoStop}%
\bibitem [{\citenamefont {Spencer}\ \emph {et~al.}(2016)\citenamefont
  {Spencer}, \citenamefont {Gajdos},\ and\ \citenamefont
  {Blumberger}}]{Spencer2016}%
  \BibitemOpen
  \bibfield  {author} {\bibinfo {author} {\bibfnamefont {J.}~\bibnamefont
  {Spencer}}, \bibinfo {author} {\bibfnamefont {F.}~\bibnamefont {Gajdos}},\
  and\ \bibinfo {author} {\bibfnamefont {J.}~\bibnamefont {Blumberger}},\
  }\bibfield  {title} {\bibinfo {title} {{FOB-SH}: {F}ragment orbital-based
  surface hopping for charge carrier transport in organic and biological
  molecules and materials},\ }\href {https://doi.org/10.1063/1.4960144}
  {\bibfield  {journal} {\bibinfo  {journal} {J. Chem. Phys.}\ }\textbf
  {\bibinfo {volume} {145}},\ \bibinfo {pages} {064102} (\bibinfo {year}
  {2016})}\BibitemShut {NoStop}%
\bibitem [{\citenamefont {Giannini}\ \emph {et~al.}(2018)\citenamefont
  {Giannini}, \citenamefont {Carof},\ and\ \citenamefont
  {Blumberger}}]{Giannini2018}%
  \BibitemOpen
  \bibfield  {author} {\bibinfo {author} {\bibfnamefont {S.}~\bibnamefont
  {Giannini}}, \bibinfo {author} {\bibfnamefont {A.}~\bibnamefont {Carof}},\
  and\ \bibinfo {author} {\bibfnamefont {J.}~\bibnamefont {Blumberger}},\
  }\bibfield  {title} {\bibinfo {title} {Crossover from hopping to band-like
  charge transport in an organic semiconductor model: {A}tomistic nonadiabatic
  molecular dynamics simulation},\ }\href
  {https://doi.org/10.1021/acs.jpclett.8b01112} {\bibfield  {journal} {\bibinfo
   {journal} {J. Phys. Chem. Lett.}\ }\textbf {\bibinfo {volume} {9}},\
  \bibinfo {pages} {3116} (\bibinfo {year} {2018})}\BibitemShut {NoStop}%
\bibitem [{\citenamefont {Giannini}\ and\ \citenamefont
  {Blumberger}(2022)}]{Giannini2022}%
  \BibitemOpen
  \bibfield  {author} {\bibinfo {author} {\bibfnamefont {S.}~\bibnamefont
  {Giannini}}\ and\ \bibinfo {author} {\bibfnamefont {J.}~\bibnamefont
  {Blumberger}},\ }\bibfield  {title} {\bibinfo {title} {Charge transport in
  organic semiconductors: {T}he perspective from nonadiabatic molecular
  dynamics},\ }\href {https://doi.org/10.1021/acs.accounts.1c00675} {\bibfield
  {journal} {\bibinfo  {journal} {Acc. Chem. Res.}\ }\textbf {\bibinfo {volume}
  {55}},\ \bibinfo {pages} {819} (\bibinfo {year} {2022})}\BibitemShut
  {NoStop}%
\bibitem [{\citenamefont {Heck}\ \emph {et~al.}(2015)\citenamefont {Heck},
  \citenamefont {Kranz}, \citenamefont {Kuba\v{r}},\ and\ \citenamefont
  {Elstner}}]{Heck2015}%
  \BibitemOpen
  \bibfield  {author} {\bibinfo {author} {\bibfnamefont {A.}~\bibnamefont
  {Heck}}, \bibinfo {author} {\bibfnamefont {J.~J.}\ \bibnamefont {Kranz}},
  \bibinfo {author} {\bibfnamefont {T.}~\bibnamefont {Kuba\v{r}}},\ and\
  \bibinfo {author} {\bibfnamefont {M.}~\bibnamefont {Elstner}},\ }\bibfield
  {title} {\bibinfo {title} {Multi-scale approach to non-adiabatic charge
  transport in high-mobility organic semiconductors},\ }\href
  {https://doi.org/10.1021/acs.jctc.5b00719} {\bibfield  {journal} {\bibinfo
  {journal} {J. Chem. Theory Comput.}\ }\textbf {\bibinfo {volume} {11}},\
  \bibinfo {pages} {5068} (\bibinfo {year} {2015})}\BibitemShut {NoStop}%
\bibitem [{\citenamefont {Heck}\ \emph {et~al.}(2016)\citenamefont {Heck},
  \citenamefont {Kranz},\ and\ \citenamefont {Elstner}}]{Heck2016}%
  \BibitemOpen
  \bibfield  {author} {\bibinfo {author} {\bibfnamefont {A.}~\bibnamefont
  {Heck}}, \bibinfo {author} {\bibfnamefont {J.~J.}\ \bibnamefont {Kranz}},\
  and\ \bibinfo {author} {\bibfnamefont {M.}~\bibnamefont {Elstner}},\
  }\bibfield  {title} {\bibinfo {title} {Simulation of temperature-dependent
  charge transport in organic semiconductors with various degrees of
  disorder},\ }\href {https://doi.org/10.1021/acs.jctc.6b00215} {\bibfield
  {journal} {\bibinfo  {journal} {J. Chem. Theory Comput.}\ }\textbf {\bibinfo
  {volume} {12}},\ \bibinfo {pages} {3087} (\bibinfo {year}
  {2016})}\BibitemShut {NoStop}%
\bibitem [{\citenamefont {Jiang}\ \emph {et~al.}(2016)\citenamefont {Jiang},
  \citenamefont {Zhong}, \citenamefont {Shi}, \citenamefont {Peng},
  \citenamefont {Geng}, \citenamefont {Zhao},\ and\ \citenamefont
  {Shuai}}]{Jiang2016}%
  \BibitemOpen
  \bibfield  {author} {\bibinfo {author} {\bibfnamefont {Y.}~\bibnamefont
  {Jiang}}, \bibinfo {author} {\bibfnamefont {X.}~\bibnamefont {Zhong}},
  \bibinfo {author} {\bibfnamefont {W.}~\bibnamefont {Shi}}, \bibinfo {author}
  {\bibfnamefont {Q.}~\bibnamefont {Peng}}, \bibinfo {author} {\bibfnamefont
  {H.}~\bibnamefont {Geng}}, \bibinfo {author} {\bibfnamefont {Y.}~\bibnamefont
  {Zhao}},\ and\ \bibinfo {author} {\bibfnamefont {Z.}~\bibnamefont {Shuai}},\
  }\bibfield  {title} {\bibinfo {title} {Nuclear quantum tunnelling and carrier
  delocalization effects to bridge the gap between hopping and bandlike
  behaviors in organic semiconductors},\ }\href
  {https://doi.org/10.1039/C5NH00054H} {\bibfield  {journal} {\bibinfo
  {journal} {Nanoscale Horiz.}\ }\textbf {\bibinfo {volume} {1}},\ \bibinfo
  {pages} {53} (\bibinfo {year} {2016})}\BibitemShut {NoStop}%
\bibitem [{\citenamefont {Fratini}\ \emph {et~al.}(2016)\citenamefont
  {Fratini}, \citenamefont {Mayou},\ and\ \citenamefont
  {Ciuchi}}]{Fratini2016}%
  \BibitemOpen
  \bibfield  {author} {\bibinfo {author} {\bibfnamefont {S.}~\bibnamefont
  {Fratini}}, \bibinfo {author} {\bibfnamefont {D.}~\bibnamefont {Mayou}},\
  and\ \bibinfo {author} {\bibfnamefont {S.}~\bibnamefont {Ciuchi}},\
  }\bibfield  {title} {\bibinfo {title} {The transient localization scenario
  for charge transport in crystalline organic materials},\ }\href
  {https://doi.org/https://doi.org/10.1002/adfm.201502386} {\bibfield
  {journal} {\bibinfo  {journal} {Adv. Funct. Mater.}\ }\textbf {\bibinfo
  {volume} {26}},\ \bibinfo {pages} {2292} (\bibinfo {year}
  {2016})}\BibitemShut {NoStop}%
\bibitem [{\citenamefont {Nematiaram}\ \emph {et~al.}(2019)\citenamefont
  {Nematiaram}, \citenamefont {Ciuchi}, \citenamefont {Xie}, \citenamefont
  {Fratini},\ and\ \citenamefont {Troisi}}]{Nematiaram2019}%
  \BibitemOpen
  \bibfield  {author} {\bibinfo {author} {\bibfnamefont {T.}~\bibnamefont
  {Nematiaram}}, \bibinfo {author} {\bibfnamefont {S.}~\bibnamefont {Ciuchi}},
  \bibinfo {author} {\bibfnamefont {X.}~\bibnamefont {Xie}}, \bibinfo {author}
  {\bibfnamefont {S.}~\bibnamefont {Fratini}},\ and\ \bibinfo {author}
  {\bibfnamefont {A.}~\bibnamefont {Troisi}},\ }\bibfield  {title} {\bibinfo
  {title} {Practical computation of the charge mobility in molecular
  semiconductors using transient localization theory},\ }\href
  {https://doi.org/10.1021/acs.jpcc.8b11916} {\bibfield  {journal} {\bibinfo
  {journal} {J. Phys. Chem. C}\ }\textbf {\bibinfo {volume} {123}},\ \bibinfo
  {pages} {6989} (\bibinfo {year} {2019})}\BibitemShut {NoStop}%
\bibitem [{\citenamefont {Varvelo}\ \emph {et~al.}(2021)\citenamefont
  {Varvelo}, \citenamefont {Lynd},\ and\ \citenamefont
  {Bennett}}]{Varvelo2021}%
  \BibitemOpen
  \bibfield  {author} {\bibinfo {author} {\bibfnamefont {L.}~\bibnamefont
  {Varvelo}}, \bibinfo {author} {\bibfnamefont {J.~K.}\ \bibnamefont {Lynd}},\
  and\ \bibinfo {author} {\bibfnamefont {D.~I.~G.}\ \bibnamefont {Bennett}},\
  }\bibfield  {title} {\bibinfo {title} {Formally exact simulations of
  mesoscale exciton dynamics in molecular materials},\ }\href
  {https://doi.org/10.1039/D1SC01448J} {\bibfield  {journal} {\bibinfo
  {journal} {Chem. Sci.}\ }\textbf {\bibinfo {volume} {12}},\ \bibinfo {pages}
  {9704} (\bibinfo {year} {2021})}\BibitemShut {NoStop}%
\bibitem [{\citenamefont {Li}\ \emph {et~al.}(2020)\citenamefont {Li},
  \citenamefont {Ren},\ and\ \citenamefont {Shuai}}]{Li2020}%
  \BibitemOpen
  \bibfield  {author} {\bibinfo {author} {\bibfnamefont {W.}~\bibnamefont
  {Li}}, \bibinfo {author} {\bibfnamefont {J.}~\bibnamefont {Ren}},\ and\
  \bibinfo {author} {\bibfnamefont {Z.}~\bibnamefont {Shuai}},\ }\bibfield
  {title} {\bibinfo {title} {Finite-temperature {TD-DMRG} for the carrier
  mobility of organic semiconductors},\ }\href
  {https://doi.org/10.1021/acs.jpclett.0c01072} {\bibfield  {journal} {\bibinfo
   {journal} {J. Phys. Chem. Lett.}\ }\textbf {\bibinfo {volume} {11}},\
  \bibinfo {pages} {4930} (\bibinfo {year} {2020})}\BibitemShut {NoStop}%
\bibitem [{\citenamefont {Li}\ \emph {et~al.}(2021)\citenamefont {Li},
  \citenamefont {Ren},\ and\ \citenamefont {Shuai}}]{Li2021}%
  \BibitemOpen
  \bibfield  {author} {\bibinfo {author} {\bibfnamefont {W.}~\bibnamefont
  {Li}}, \bibinfo {author} {\bibfnamefont {J.}~\bibnamefont {Ren}},\ and\
  \bibinfo {author} {\bibfnamefont {Z.}~\bibnamefont {Shuai}},\ }\bibfield
  {title} {\bibinfo {title} {A general charge transport picture for organic
  semiconductors with nonlocal electron-phonon couplings},\ }\href
  {https://doi.org/10.1038/s41467-021-24520-y} {\bibfield  {journal} {\bibinfo
  {journal} {Nat. Commun.}\ }\textbf {\bibinfo {volume} {12}},\ \bibinfo
  {pages} {4260} (\bibinfo {year} {2021})}\BibitemShut {NoStop}%
\bibitem [{\citenamefont {Savoie}\ \emph {et~al.}(2014)\citenamefont {Savoie},
  \citenamefont {Kohlstedt}, \citenamefont {Jackson}, \citenamefont {Chen},
  \citenamefont {de~la Cruz}, \citenamefont {Schatz}, \citenamefont {Marks},\
  and\ \citenamefont {Ratner}}]{Savoie2014}%
  \BibitemOpen
  \bibfield  {author} {\bibinfo {author} {\bibfnamefont {B.~M.}\ \bibnamefont
  {Savoie}}, \bibinfo {author} {\bibfnamefont {K.~L.}\ \bibnamefont
  {Kohlstedt}}, \bibinfo {author} {\bibfnamefont {N.~E.}\ \bibnamefont
  {Jackson}}, \bibinfo {author} {\bibfnamefont {L.~X.}\ \bibnamefont {Chen}},
  \bibinfo {author} {\bibfnamefont {M.~O.}\ \bibnamefont {de~la Cruz}},
  \bibinfo {author} {\bibfnamefont {G.~C.}\ \bibnamefont {Schatz}}, \bibinfo
  {author} {\bibfnamefont {T.~J.}\ \bibnamefont {Marks}},\ and\ \bibinfo
  {author} {\bibfnamefont {M.~A.}\ \bibnamefont {Ratner}},\ }\bibfield  {title}
  {\bibinfo {title} {Mesoscale molecular network formation in amorphous organic
  materials},\ }\href {https://doi.org/10.1073/pnas.1409514111} {\bibfield
  {journal} {\bibinfo  {journal} {Proc. Natl. Acad. Sci. U.S.A.}\ }\textbf
  {\bibinfo {volume} {111}},\ \bibinfo {pages} {10055} (\bibinfo {year}
  {2014})}\BibitemShut {NoStop}%
\bibitem [{\citenamefont {Jackson}\ \emph {et~al.}(2016)\citenamefont
  {Jackson}, \citenamefont {Chen},\ and\ \citenamefont {Ratner}}]{Jackson2016}%
  \BibitemOpen
  \bibfield  {author} {\bibinfo {author} {\bibfnamefont {N.~E.}\ \bibnamefont
  {Jackson}}, \bibinfo {author} {\bibfnamefont {L.~X.}\ \bibnamefont {Chen}},\
  and\ \bibinfo {author} {\bibfnamefont {M.~A.}\ \bibnamefont {Ratner}},\
  }\bibfield  {title} {\bibinfo {title} {Charge transport network dynamics in
  molecular aggregates},\ }\href {https://doi.org/10.1073/pnas.1601915113}
  {\bibfield  {journal} {\bibinfo  {journal} {Proc. Natl. Acad. Sci. U.S.A.}\
  }\textbf {\bibinfo {volume} {113}},\ \bibinfo {pages} {8595} (\bibinfo {year}
  {2016})}\BibitemShut {NoStop}%
\bibitem [{\citenamefont {Jankovi\'c}\ and\ \citenamefont
  {Vukmirovi\'c}(2020)}]{Jankovic2020}%
  \BibitemOpen
  \bibfield  {author} {\bibinfo {author} {\bibfnamefont {V.}~\bibnamefont
  {Jankovi\'c}}\ and\ \bibinfo {author} {\bibfnamefont {N.}~\bibnamefont
  {Vukmirovi\'c}},\ }\bibfield  {title} {\bibinfo {title} {Energy-temporal
  pathways of free-charge formation at organic bilayers: {C}ompetition of
  delocalization, disorder, and polaronic effects},\ }\href
  {https://doi.org/10.1021/acs.jpcc.9b10862} {\bibfield  {journal} {\bibinfo
  {journal} {J. Phys. Chem. C}\ }\textbf {\bibinfo {volume} {124}},\ \bibinfo
  {pages} {4378} (\bibinfo {year} {2020})}\BibitemShut {NoStop}%
\bibitem [{\citenamefont {Jang}(2011)}]{Jang2011}%
  \BibitemOpen
  \bibfield  {author} {\bibinfo {author} {\bibfnamefont {S.}~\bibnamefont
  {Jang}},\ }\bibfield  {title} {\bibinfo {title} {Theory of multichromophoric
  coherent resonance energy transfer: {A} polaronic quantum master equation
  approach},\ }\href {https://doi.org/10.1063/1.3608914} {\bibfield  {journal}
  {\bibinfo  {journal} {J. Chem. Phys.}\ }\textbf {\bibinfo {volume} {135}},\
  \bibinfo {pages} {034105} (\bibinfo {year} {2011})}\BibitemShut {NoStop}%
\bibitem [{\citenamefont {Lee}\ \emph {et~al.}(2015)\citenamefont {Lee},
  \citenamefont {Moix},\ and\ \citenamefont {Cao}}]{Lee2015}%
  \BibitemOpen
  \bibfield  {author} {\bibinfo {author} {\bibfnamefont {C.~K.}\ \bibnamefont
  {Lee}}, \bibinfo {author} {\bibfnamefont {J.}~\bibnamefont {Moix}},\ and\
  \bibinfo {author} {\bibfnamefont {J.}~\bibnamefont {Cao}},\ }\bibfield
  {title} {\bibinfo {title} {Coherent quantum transport in disordered systems:
  {A} unified polaron treatment of hopping and band-like transport},\ }\href
  {https://doi.org/10.1063/1.4918736} {\bibfield  {journal} {\bibinfo
  {journal} {J. Chem. Phys.}\ }\textbf {\bibinfo {volume} {142}},\ \bibinfo
  {pages} {164103} (\bibinfo {year} {2015})}\BibitemShut {NoStop}%
\bibitem [{\citenamefont {Balzer}\ and\ \citenamefont
  {Kassal}(2022)}]{Balzer2022}%
  \BibitemOpen
  \bibfield  {author} {\bibinfo {author} {\bibfnamefont {D.}~\bibnamefont
  {Balzer}}\ and\ \bibinfo {author} {\bibfnamefont {I.}~\bibnamefont
  {Kassal}},\ }\bibfield  {title} {\bibinfo {title} {Even a little
  delocalization produces large kinetic enhancements of charge-separation
  efficiency in organic photovoltaics},\ }\href
  {https://doi.org/10.1126/sciadv.abl9692} {\bibfield  {journal} {\bibinfo
  {journal} {Sci. Adv.}\ }\textbf {\bibinfo {volume} {8}},\ \bibinfo {pages}
  {eabl9692} (\bibinfo {year} {2022})}\BibitemShut {NoStop}%
\bibitem [{\citenamefont {B\"assler}(1993)}]{Bassler1993}%
  \BibitemOpen
  \bibfield  {author} {\bibinfo {author} {\bibfnamefont {H.}~\bibnamefont
  {B\"assler}},\ }\bibfield  {title} {\bibinfo {title} {Charge transport in
  disordered organic photoconductors. {A} {M}onte {C}arlo simulation study},\
  }\href {https://doi.org/10.1002/pssb.2221750102} {\bibfield  {journal}
  {\bibinfo  {journal} {Phys. Status Solidi B}\ }\textbf {\bibinfo {volume}
  {175}},\ \bibinfo {pages} {15} (\bibinfo {year} {1993})}\BibitemShut
  {NoStop}%
\bibitem [{\citenamefont {Marcus}(1956)}]{Marcus1956}%
  \BibitemOpen
  \bibfield  {author} {\bibinfo {author} {\bibfnamefont {R.~A.}\ \bibnamefont
  {Marcus}},\ }\bibfield  {title} {\bibinfo {title} {On the theory of
  oxidation‐reduction reactions involving electron transfer. {I}},\ }\href
  {https://doi.org/10.1063/1.1742723} {\bibfield  {journal} {\bibinfo
  {journal} {J. Chem. Phys.}\ }\textbf {\bibinfo {volume} {24}},\ \bibinfo
  {pages} {966} (\bibinfo {year} {1956})}\BibitemShut {NoStop}%
\bibitem [{\citenamefont {Miller}\ and\ \citenamefont
  {Abrahams}(1960)}]{Miller1960}%
  \BibitemOpen
  \bibfield  {author} {\bibinfo {author} {\bibfnamefont {A.}~\bibnamefont
  {Miller}}\ and\ \bibinfo {author} {\bibfnamefont {E.}~\bibnamefont
  {Abrahams}},\ }\bibfield  {title} {\bibinfo {title} {Impurity conduction at
  low concentrations},\ }\href {https://doi.org/10.1103/physrev.120.745}
  {\bibfield  {journal} {\bibinfo  {journal} {Phys. Rev.}\ }\textbf {\bibinfo
  {volume} {120}},\ \bibinfo {pages} {745} (\bibinfo {year}
  {1960})}\BibitemShut {NoStop}%
\bibitem [{\citenamefont {Baranovskii}(2014)}]{Baranovskii2014}%
  \BibitemOpen
  \bibfield  {author} {\bibinfo {author} {\bibfnamefont {S.~D.}\ \bibnamefont
  {Baranovskii}},\ }\bibfield  {title} {\bibinfo {title} {Theoretical
  description of charge transport in disordered organic semiconductors},\
  }\href {https://doi.org/10.1002/pssb.201350339} {\bibfield  {journal}
  {\bibinfo  {journal} {Phys. Status Solidi B}\ }\textbf {\bibinfo {volume}
  {251}},\ \bibinfo {pages} {487} (\bibinfo {year} {2014})}\BibitemShut
  {NoStop}%
\bibitem [{\citenamefont {Wilken}\ \emph {et~al.}(2020)\citenamefont {Wilken},
  \citenamefont {Upreti}, \citenamefont {Melianas}, \citenamefont
  {Dahlstr\"om}, \citenamefont {Persson}, \citenamefont {Olsson}, \citenamefont
  {\"Osterbacka},\ and\ \citenamefont {Kemerink}}]{Wilken2020}%
  \BibitemOpen
  \bibfield  {author} {\bibinfo {author} {\bibfnamefont {S.}~\bibnamefont
  {Wilken}}, \bibinfo {author} {\bibfnamefont {T.}~\bibnamefont {Upreti}},
  \bibinfo {author} {\bibfnamefont {A.}~\bibnamefont {Melianas}}, \bibinfo
  {author} {\bibfnamefont {S.}~\bibnamefont {Dahlstr\"om}}, \bibinfo {author}
  {\bibfnamefont {G.}~\bibnamefont {Persson}}, \bibinfo {author} {\bibfnamefont
  {E.}~\bibnamefont {Olsson}}, \bibinfo {author} {\bibfnamefont
  {R.}~\bibnamefont {\"Osterbacka}},\ and\ \bibinfo {author} {\bibfnamefont
  {M.}~\bibnamefont {Kemerink}},\ }\bibfield  {title} {\bibinfo {title}
  {Experimentally calibrated kinetic {M}onte {C}arlo model reproduces organic
  solar cell current–voltage curve},\ }\href
  {https://doi.org/https://doi.org/10.1002/solr.202000029} {\bibfield
  {journal} {\bibinfo  {journal} {Sol. RRL}\ }\textbf {\bibinfo {volume} {4}},\
  \bibinfo {pages} {2000029} (\bibinfo {year} {2020})}\BibitemShut {NoStop}%
\bibitem [{\citenamefont {Zojer}(2021)}]{Zojer2021}%
  \BibitemOpen
  \bibfield  {author} {\bibinfo {author} {\bibfnamefont {K.}~\bibnamefont
  {Zojer}},\ }\bibfield  {title} {\bibinfo {title} {Simulation of charge
  carriers in organic electronic devices: {M}ethods with their fundamentals and
  applications},\ }\href {https://doi.org/10.1002/adom.202100219} {\bibfield
  {journal} {\bibinfo  {journal} {Adv. Opt. Mater.}\ }\textbf {\bibinfo
  {volume} {9}},\ \bibinfo {pages} {2100219} (\bibinfo {year}
  {2021})}\BibitemShut {NoStop}%
\bibitem [{\citenamefont {Upreti}\ \emph {et~al.}(2021)\citenamefont {Upreti},
  \citenamefont {Wilken}, \citenamefont {Zhang},\ and\ \citenamefont
  {Kemerink}}]{Upreti2021}%
  \BibitemOpen
  \bibfield  {author} {\bibinfo {author} {\bibfnamefont {T.}~\bibnamefont
  {Upreti}}, \bibinfo {author} {\bibfnamefont {S.}~\bibnamefont {Wilken}},
  \bibinfo {author} {\bibfnamefont {H.}~\bibnamefont {Zhang}},\ and\ \bibinfo
  {author} {\bibfnamefont {M.}~\bibnamefont {Kemerink}},\ }\bibfield  {title}
  {\bibinfo {title} {Slow relaxation of photogenerated charge carriers boosts
  open-circuit voltage of organic solar cells},\ }\href
  {https://doi.org/10.1021/acs.jpclett.1c02235} {\bibfield  {journal} {\bibinfo
   {journal} {J. Phys. Chem. Lett.}\ }\textbf {\bibinfo {volume} {12}},\
  \bibinfo {pages} {9874} (\bibinfo {year} {2021})}\BibitemShut {NoStop}%
\bibitem [{\citenamefont {Vukmirovi\'c}\ and\ \citenamefont
  {Wang}(2010)}]{Vukmirovic2010}%
  \BibitemOpen
  \bibfield  {author} {\bibinfo {author} {\bibfnamefont {N.}~\bibnamefont
  {Vukmirovi\'c}}\ and\ \bibinfo {author} {\bibfnamefont {L.-W.}\ \bibnamefont
  {Wang}},\ }\bibfield  {title} {\bibinfo {title} {Carrier hopping in
  disordered semiconducting polymers: {H}ow accurate is the
  {M}iller–{A}brahams model?},\ }\href {https://doi.org/10.1063/1.3474618}
  {\bibfield  {journal} {\bibinfo  {journal} {Appl. Phys. Lett.}\ }\textbf
  {\bibinfo {volume} {97}},\ \bibinfo {pages} {043305} (\bibinfo {year}
  {2010})}\BibitemShut {NoStop}%
\bibitem [{\citenamefont {Br\'{e}das}\ \emph {et~al.}(2002)\citenamefont
  {Br\'{e}das}, \citenamefont {Calbert}, \citenamefont {da~Silva~Filho},\ and\
  \citenamefont {Cornil}}]{Bredas2002}%
  \BibitemOpen
  \bibfield  {author} {\bibinfo {author} {\bibfnamefont {J.~L.}\ \bibnamefont
  {Br\'{e}das}}, \bibinfo {author} {\bibfnamefont {J.~P.}\ \bibnamefont
  {Calbert}}, \bibinfo {author} {\bibfnamefont {D.~A.}\ \bibnamefont
  {da~Silva~Filho}},\ and\ \bibinfo {author} {\bibfnamefont {J.}~\bibnamefont
  {Cornil}},\ }\bibfield  {title} {\bibinfo {title} {Organic semiconductors:
  {A} theoretical characterization of the basic parameters governing charge
  transport},\ }\href {https://doi.org/10.1073/pnas.092143399} {\bibfield
  {journal} {\bibinfo  {journal} {Proc. Natl. Acad. Sci. U.S.A.}\ }\textbf
  {\bibinfo {volume} {99}},\ \bibinfo {pages} {5804} (\bibinfo {year}
  {2002})}\BibitemShut {NoStop}%
\bibitem [{\citenamefont {Hoffmann}\ \emph {et~al.}(2012)\citenamefont
  {Hoffmann}, \citenamefont {Athanasopoulos}, \citenamefont {Beljonne},
  \citenamefont {B\"assler},\ and\ \citenamefont {K\"ohler}}]{Hoffmann2012}%
  \BibitemOpen
  \bibfield  {author} {\bibinfo {author} {\bibfnamefont {S.~T.}\ \bibnamefont
  {Hoffmann}}, \bibinfo {author} {\bibfnamefont {S.}~\bibnamefont
  {Athanasopoulos}}, \bibinfo {author} {\bibfnamefont {D.}~\bibnamefont
  {Beljonne}}, \bibinfo {author} {\bibfnamefont {H.}~\bibnamefont
  {B\"assler}},\ and\ \bibinfo {author} {\bibfnamefont {A.}~\bibnamefont
  {K\"ohler}},\ }\bibfield  {title} {\bibinfo {title} {How do triplets and
  charges move in disordered organic semiconductors? {A} {M}onte {C}arlo study
  comprising the equilibrium and nonequilibrium regime},\ }\href
  {https://doi.org/10.1021/jp305062p} {\bibfield  {journal} {\bibinfo
  {journal} {J. Phys. Chem. C}\ }\textbf {\bibinfo {volume} {116}},\ \bibinfo
  {pages} {16371} (\bibinfo {year} {2012})}\BibitemShut {NoStop}%
\bibitem [{\citenamefont {Shuai}\ \emph {et~al.}(2020)\citenamefont {Shuai},
  \citenamefont {Li}, \citenamefont {Ren}, \citenamefont {Jiang},\ and\
  \citenamefont {Geng}}]{Shuai2020}%
  \BibitemOpen
  \bibfield  {author} {\bibinfo {author} {\bibfnamefont {Z.}~\bibnamefont
  {Shuai}}, \bibinfo {author} {\bibfnamefont {W.}~\bibnamefont {Li}}, \bibinfo
  {author} {\bibfnamefont {J.}~\bibnamefont {Ren}}, \bibinfo {author}
  {\bibfnamefont {Y.}~\bibnamefont {Jiang}},\ and\ \bibinfo {author}
  {\bibfnamefont {H.}~\bibnamefont {Geng}},\ }\bibfield  {title} {\bibinfo
  {title} {Applying {M}arcus theory to describe the carrier transports in
  organic semiconductors: {L}imitations and beyond},\ }\href
  {https://doi.org/10.1063/5.0018312} {\bibfield  {journal} {\bibinfo
  {journal} {J. Chem. Phys.}\ }\textbf {\bibinfo {volume} {153}},\ \bibinfo
  {pages} {080902} (\bibinfo {year} {2020})}\BibitemShut {NoStop}%
\bibitem [{\citenamefont {Melianas}\ \emph {et~al.}(2014)\citenamefont
  {Melianas}, \citenamefont {Pranculis}, \citenamefont {Devi\v{z}is},
  \citenamefont {Gulbinas}, \citenamefont {Ingan\"as},\ and\ \citenamefont
  {Kemerink}}]{Melianas2014}%
  \BibitemOpen
  \bibfield  {author} {\bibinfo {author} {\bibfnamefont {A.}~\bibnamefont
  {Melianas}}, \bibinfo {author} {\bibfnamefont {V.}~\bibnamefont {Pranculis}},
  \bibinfo {author} {\bibfnamefont {A.}~\bibnamefont {Devi\v{z}is}}, \bibinfo
  {author} {\bibfnamefont {V.}~\bibnamefont {Gulbinas}}, \bibinfo {author}
  {\bibfnamefont {O.}~\bibnamefont {Ingan\"as}},\ and\ \bibinfo {author}
  {\bibfnamefont {M.}~\bibnamefont {Kemerink}},\ }\bibfield  {title} {\bibinfo
  {title} {Dispersion‐dominated photocurrent in polymer:fullerene solar
  cells},\ }\href {https://doi.org/10.1002/adfm.201400404} {\bibfield
  {journal} {\bibinfo  {journal} {Adv. Funct. Mater.}\ }\textbf {\bibinfo
  {volume} {24}},\ \bibinfo {pages} {4507} (\bibinfo {year}
  {2014})}\BibitemShut {NoStop}%
\bibitem [{\citenamefont {Melianas}\ and\ \citenamefont
  {Kemerink}(2019)}]{Melianas2019}%
  \BibitemOpen
  \bibfield  {author} {\bibinfo {author} {\bibfnamefont {A.}~\bibnamefont
  {Melianas}}\ and\ \bibinfo {author} {\bibfnamefont {M.}~\bibnamefont
  {Kemerink}},\ }\bibfield  {title} {\bibinfo {title} {Photogenerated charge
  transport in organic electronic materials: {E}xperiments confirmed by
  simulations},\ }\href
  {https://doi.org/https://doi.org/10.1002/adma.201806004} {\bibfield
  {journal} {\bibinfo  {journal} {Adv. Mater.}\ }\textbf {\bibinfo {volume}
  {31}},\ \bibinfo {pages} {1806004} (\bibinfo {year} {2019})}\BibitemShut
  {NoStop}%
\bibitem [{\citenamefont {Vissenberg}\ and\ \citenamefont
  {Matters}(1998)}]{Vissenberg1998}%
  \BibitemOpen
  \bibfield  {author} {\bibinfo {author} {\bibfnamefont {M.~C. J.~M.}\
  \bibnamefont {Vissenberg}}\ and\ \bibinfo {author} {\bibfnamefont
  {M.}~\bibnamefont {Matters}},\ }\bibfield  {title} {\bibinfo {title} {Theory
  of the field-effect mobility in amorphous organic transistors},\ }\href
  {https://doi.org/10.1103/PhysRevB.57.12964} {\bibfield  {journal} {\bibinfo
  {journal} {Phys. Rev. B}\ }\textbf {\bibinfo {volume} {57}},\ \bibinfo
  {pages} {12964} (\bibinfo {year} {1998})}\BibitemShut {NoStop}%
\bibitem [{\citenamefont {Gartstein}\ and\ \citenamefont
  {Conwell}(1995)}]{Gartstein1995}%
  \BibitemOpen
  \bibfield  {author} {\bibinfo {author} {\bibfnamefont {Y.}~\bibnamefont
  {Gartstein}}\ and\ \bibinfo {author} {\bibfnamefont {E.}~\bibnamefont
  {Conwell}},\ }\bibfield  {title} {\bibinfo {title} {High-field hopping
  mobility in molecular systems with spatially correlated energetic disorder},\
  }\href {https://doi.org/https://doi.org/10.1016/0009-2614(95)01031-4}
  {\bibfield  {journal} {\bibinfo  {journal} {Chem. Phys. Lett.}\ }\textbf
  {\bibinfo {volume} {245}},\ \bibinfo {pages} {351} (\bibinfo {year}
  {1995})}\BibitemShut {NoStop}%
\bibitem [{\citenamefont {Troisi}\ and\ \citenamefont
  {Orlandi}(2006{\natexlab{a}})}]{Troisi2006}%
  \BibitemOpen
  \bibfield  {author} {\bibinfo {author} {\bibfnamefont {A.}~\bibnamefont
  {Troisi}}\ and\ \bibinfo {author} {\bibfnamefont {G.}~\bibnamefont
  {Orlandi}},\ }\bibfield  {title} {\bibinfo {title} {Charge-transport regime
  of crystalline organic semiconductors: {D}iffusion limited by thermal
  off-diagonal electronic disorder},\ }\href
  {https://doi.org/10.1103/PhysRevLett.96.086601} {\bibfield  {journal}
  {\bibinfo  {journal} {Phys. Rev. Lett.}\ }\textbf {\bibinfo {volume} {96}},\
  \bibinfo {pages} {086601} (\bibinfo {year} {2006}{\natexlab{a}})}\BibitemShut
  {NoStop}%
\bibitem [{\citenamefont {Troisi}\ and\ \citenamefont
  {Orlandi}(2006{\natexlab{b}})}]{Troisi2006_2}%
  \BibitemOpen
  \bibfield  {author} {\bibinfo {author} {\bibfnamefont {A.}~\bibnamefont
  {Troisi}}\ and\ \bibinfo {author} {\bibfnamefont {G.}~\bibnamefont
  {Orlandi}},\ }\bibfield  {title} {\bibinfo {title} {Dynamics of the
  intermolecular transfer integral in crystalline organic semiconductors},\
  }\href {https://doi.org/10.1021/jp055432g} {\bibfield  {journal} {\bibinfo
  {journal} {J. Phys. Chem. A}\ }\textbf {\bibinfo {volume} {110}},\ \bibinfo
  {pages} {4065} (\bibinfo {year} {2006}{\natexlab{b}})}\BibitemShut {NoStop}%
\end{thebibliography}%

\section*{Appendices} 

\appendix

\setcounter{section}{0}
\renewcommand{\thesection}{A\arabic{section}}%
\setcounter{equation}{0}
\renewcommand{\theequation}{A\arabic{equation}}%
\setcounter{figure}{0}
\renewcommand{\thefigure}{A\arabic{figure}}%

\section{jKMC rate derivation}
\label{app:jKMC_rate_derivation}

\begin{figure}
    \centering
    \includegraphics[width=\columnwidth]{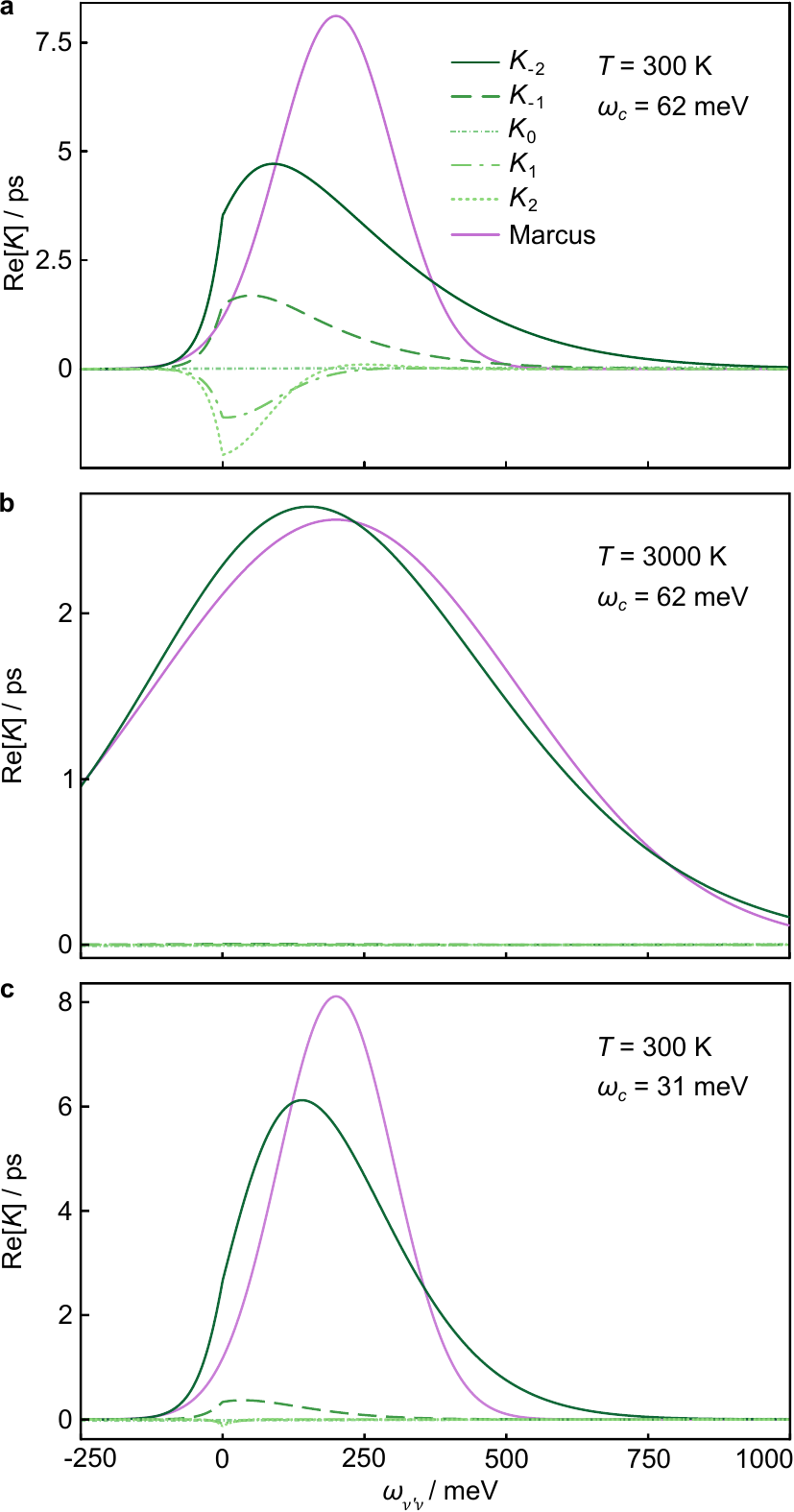}
    \caption{\textbf{Comparison of different $K_{\Delta(ij,i'j')}(\omega)$}, shown as functions of the polaron energy difference $\omega_{\nu\nu'}$. Results are calculated for $\lambda=\SI{200}{meV}$ and \textbf{(a)} the parameters in this work ($T=\SI{300}{K}$, $\omega_c=\SI{62}{meV}$), \textbf{(b)} at higher temperature ($T=\SI{3000}{K}$), and \textbf{(c)} at lower cutoff frequency ($\omega_c=\SI{31}{meV}$). At high temperatures or low cutoff frequencies, four of the five $K_{\Delta(ij,i'j')}(\omega)$ become negligible, and the surviving $K_{-2}(\omega)$ approaches $k_\mathrm{Marcus}(\omega)/2J^2$ (shown in purple).}
    \label{fig:K_comparison}
\end{figure}

The foundation of jumping kinetic Monte Carlo (jKMC) is the secular polaron transformed Redfield equation (sPTRE)~\cite{Jang2011,Lee2015}, which provides the hopping rate from any polaron $\nu$ to polaron $\nu'$ as
\begin{multline}
    \label{appeqn:sPTRE}
     R_{\nu\nu'}=\sum_{\langle i,j\rangle, \langle i',j'\rangle} 2J^2\mathrm{Re}\big[ \braket{\nu|i}\braket{j|\nu'}\braket{\nu'|i'}\braket{j'|\nu}\\[-0.3cm]\times K_{\Delta(ij,i'j')}\left(\omega_{\nu\nu'}\right)\big],
\end{multline}
where $\langle i,j\rangle$ and $\langle i',j'\rangle$ are nearest-neighbour pairs of sites, $J$ is the nearest-neighbour electronic coupling, $\omega_{\nu\nu'}=E_{\nu}-E_{\nu'}$ is the energy difference between polarons, and 
\begin{equation}
    \label{appeqn:Fourier_transformed_BCF}
    K_{\Delta(ij,i'j')}(\omega)=\int_0^{\infty}d\tau \, e^{i\omega\tau}\langle \hat{V}_{ij}(\tau)\hat{V}_{i'j'}(0) \rangle_{B} 
\end{equation}
is the half-range Fourier transform of the bath correlation function
\begin{equation}
    \label{appeqn:Bath_correlation_function}
    \langle \hat{V}_{ij}(\tau)\hat{V}_{i'j'}(0) \rangle_{B}=\kappa^2 (e^{-\Delta(ij,i'j')\phi(\tau)}-1), 
\end{equation}
where $\Delta(ij,i'j')=\delta_{ii'}-\delta_{ij'}+\delta_{jj'}-\delta_{ji'}$,
\begin{equation}
    \label{appeqn:Kappa}
    \kappa=\exp\left(-\int_0^{\infty}\frac{d\omega}{\pi} \frac{J(\omega)}{\omega^2}\coth\left(\frac{\omega}{2k_\mathrm{B}T}\right)\right),
\end{equation} and
\begin{equation}
    \label{appeqn:Phi}
    \phi(\tau)=\int_0^{\infty} \frac{d\omega}{\pi} \frac{J(\omega)}{\omega^2}\bigg(\!\cos(\omega\tau)\coth\bigg(\frac{\omega}{2k_\mathrm{B}T}\bigg)-i\sin(\omega\tau)\bigg),
\end{equation}
and we assume a super-Ohmic spectral density $J(\omega)=\frac{\pi\lambda}{4}(\omega/\omega_c)^3e^{-\omega/\omega_c}$ with cutoff frequency $\omega_c=\SI{62}{meV}$~\cite{Lee2015} and reorganisation energy $\lambda$.

We assume that the polaron wavefunctions are identical, spherically symmetric, and exponentially decaying in the site basis,
\begin{equation}
    \label{appeqn:Spherical_polaron_approximation}
    \ket{\nu}=A \sum_i  \exp\left(-\frac{d_{i \nu}}{r_\mathrm{deloc}}\right)\ket{i},
\end{equation}
where $d_{i \nu}$ is the distance between the centre of the spherical polaron $\nu$ and site $i$, $r_\mathrm{deloc}$ is the delocalisation radius that characterises the size of the wavefunction, and the normalisation prefactor is $A=\left(\sum_i \exp(-2d_{i \nu}/r_\mathrm{deloc})\right)^{-1/2}.$
Substituting the spherical-polaron approximation into the sPTRE yields
\begin{multline}
    \label{appeqn:jKMC_allK}
    R_{\nu\nu'}=\sum_{\langle i,j\rangle, \langle i',j'\rangle} 2J^2A^4 \mathrm{Re}\left[K_{\Delta(ij,i'j')}\left(\omega_{\nu\nu'}\right)\right]\\ \times \exp\left(-\frac{d_{i\nu}+d_{j\nu'}+d_{i'\nu'}+d_{j'\nu}}{r_\mathrm{deloc}}\right).
\end{multline}

\begin{figure}
    \centering
    \includegraphics[width=\columnwidth]{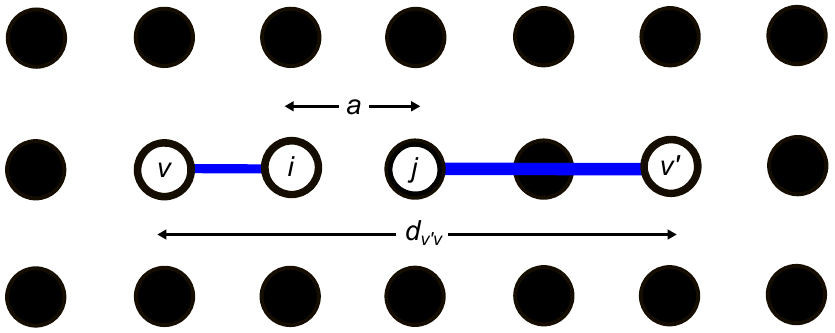}
    \caption{\textbf{The arrangement of sites whose contribution dominates the jKMC rate.} The dominant terms in the sPTRE minimise the distance $d_{i\nu}+d_{j\nu'}$ (shown in blue). On the cubic lattice, the simplest case is if the initial and final states $\nu$ and $\nu'$ lie in the same row (or column) of the lattice, as shown above. In that case, there are $d_{\nu\nu'}/a$ positions for the nearest-neighbours $i$ and $j$ which minimise this distance to $d_{\nu\nu'}-a$, and which occur when all four points are collinear. 
    We use the same result ($d_{\nu\nu'}/a$ positions with total distance $d_{\nu\nu'}-a$) even if $\nu$ and $\nu'$ do not lie in the same row or column. This approximation is justified because the particular shape of the lattice is not an essential part of the jKMC model.}
    \label{fig:Simplified_jKMC}
\end{figure}

To simplify this expression, we assume the high-temperature limit ($k_\mathrm{B}T\gg \omega_c$) in \cref{appeqn:Phi}, using $\coth(\omega/2k_\mathrm{B}T)\approx 2k_\mathrm{B}T/\omega$ to obtain
\begin{equation}
    \label{appeqn:Phi_highT}
    \phi(\tau)= \int_0^\infty \frac{d\omega}{\pi} \ \frac{J(\omega)}{\omega^2}\left(\frac{2k_\mathrm{B}T}{\omega}\cos(\omega\tau)-i\sin(\omega\tau)\right).
\end{equation}
$\phi(\tau)$ enters the integral in \cref{appeqn:Fourier_transformed_BCF} through an exponential, meaning that the integral will be dominated by contributions where $\phi(\tau)$ achieves the maximum real values. This maximisation occurs when $\cos(\omega\tau)\approx1$, i.e., when $\omega\tau\ll1$. Therefore, we take the Taylor expansions of $\cos(\omega\tau)$ and $\sin(\omega\tau)$ in \cref{appeqn:Phi_highT} to yield
\begin{equation}
    \phi(\tau)= 2k_\mathrm{B}T x_3 -(k_\mathrm{B}T\tau^2+i\tau)x_1,
\end{equation}
where we have written
\begin{equation}
    x_n = \int_0^\infty \frac{d\omega}{\pi} \ \frac{J(\omega)}{\omega^n}
\end{equation}
Similarly, the high-temperature limit of \cref{appeqn:Kappa} is
\begin{equation}
    \kappa^2=e^{-4k_\mathrm{B}T x_3}, \label{appeqn:Kappa_highT}
\end{equation}
which is exponentially small at high $T$. 

Therefore, in the high-temperature limit, the bath correlation function becomes 
\begin{multline}
    \langle \hat{V}_{ij}(\tau)\hat{V}_{i'j'}(0) \rangle_{B} = e^{-4k_\mathrm{B}T x_3} \\
    \times\left(e^{-\Delta(ij,i'j')\left(2k_\mathrm{B}T x_3 - (k_\mathrm{B}T\tau^2+i\tau) x_1 \right)}-1\right),
\end{multline}
where the coefficient $\Delta(ij,i'j')$ assumes integer values between $-2$ and 2. Of these, the $\Delta(ij,i'j')=-2$ contributions (i.e., $i=j'$ and $j=i'$) dwarf the others, and are the only ones negative and large enough to ensure that the $\kappa^2$ prefactor does not make the entire expression exponentially small in $T$. In the high-temperature limit, the other four possibilities of $\Delta(ij,i'j')$ give negligible results, as shown in \cref{fig:K_comparison}. Keeping only the $\Delta(ij,i'j')=-2$ terms in \cref{appeqn:jKMC_allK} yields 
\begin{multline}    
    \label{appeqn:jKMC_rate_Km2}
    R_{\nu\nu'}=2J^2 A^4 \mathrm{Re}\left[K_{-2}\left(\omega_{\nu\nu'}\right)\right]\\\times \sum_{\langle i,j\rangle} \exp\left(-\frac{2(d_{i\nu}+d_{j\nu'})}{r_\mathrm{deloc}}\right),
\end{multline}
where
\begin{equation}
    K_{-2}(\omega)=\int_0^\infty\! d\tau \, e^{i\omega\tau}\left(e^{-2k_\mathrm{B}T\tau^2 x_1-2i\tau x_1}-e^{-4 k_\mathrm{B}Tx_3}\right).
\end{equation}
This expression can be further simplified by neglecting the last term (which is small at high $T$), by noting that $2x_1=\lambda$ (the definition of reorganisation energy), and by extending the lower limit of the integral to $-\infty$ (because the real part of the integrand is even) to give
\begin{align}
    \mathrm{Re}\left[K_{-2}(\omega)\right]&=\frac{1}{2}\int_{-\infty}^{\infty} d\tau \ e^{i(\omega-\lambda)\tau} e^{-\lambda k_\mathrm{B}T\tau^2}\nonumber\\
    &=\sqrt{\frac{\pi}{4\lambda k_\mathrm{B}T}}\exp\left(-\frac{(\omega-\lambda)^2}{4\lambda k_\mathrm{B}T}\right)\nonumber\\
    &=\frac{k_\mathrm{Marcus}(\omega)}{2J^2},
\end{align}
where $k_\mathrm{Marcus}$ is the Marcus hopping rate.
Substituting this expression for $K_{-2}(\omega_{\nu\nu'})$ into \cref{appeqn:jKMC_rate_Km2} gives \cref{eqn:jKMC_rate_Marcus} in the main text.

\section{Simplified jKMC rate derivation}
\label{app:Simplified_jKMC_rate_derivation}
The delocalisation correction $\xi_{\nu\nu'}$ for hopping between polarons $\nu$~and~$\nu'$ can be simplified in the limit of low delocalisation to yield the simplified jKMC rate. As described in the main text, 
\begin{equation}
    \label{appeqn:Delocalisation_correction}
    \xi_{\nu\nu'}=A^4\sum_{\langle i,j\rangle} \exp\left(-\frac{2(d_{i\nu}+d_{j\nu'})}{r_\mathrm{deloc}}\right).
\end{equation}
For small $r_\mathrm{deloc}$, the sum of exponentials in \cref{appeqn:Delocalisation_correction} is dominated by the terms that minimise the distance $d_{i\nu}+d_{j\nu'}$, where $i$ and $j$ are nearest neighbours. As shown in \cref{fig:Simplified_jKMC}, there are $d_{\nu\nu'}/a$ dominant terms with the minimal $d_{i\nu}+d_{j\nu'}=d_{\nu\nu'}-a$. Because $A=1$ in the limit of localised charges, we obtain the simplified jKMC delocalisation correction
\begin{equation}
    \label{appeqn:Simplified_jKMC_rate}
    \xi_{\nu\nu'}^\mathrm{Simplified}=\frac{d_{\nu\nu'}}{a}\exp\left(-\frac{2\left(d_{\nu\nu'}-a\right)}{r_\mathrm{deloc}}\right).
\end{equation}

\section{Simplified jKMC results}
\label{app:Simplified_jKMC_results}
\begin{figure*}
    \centering
    \includegraphics[width=\textwidth]{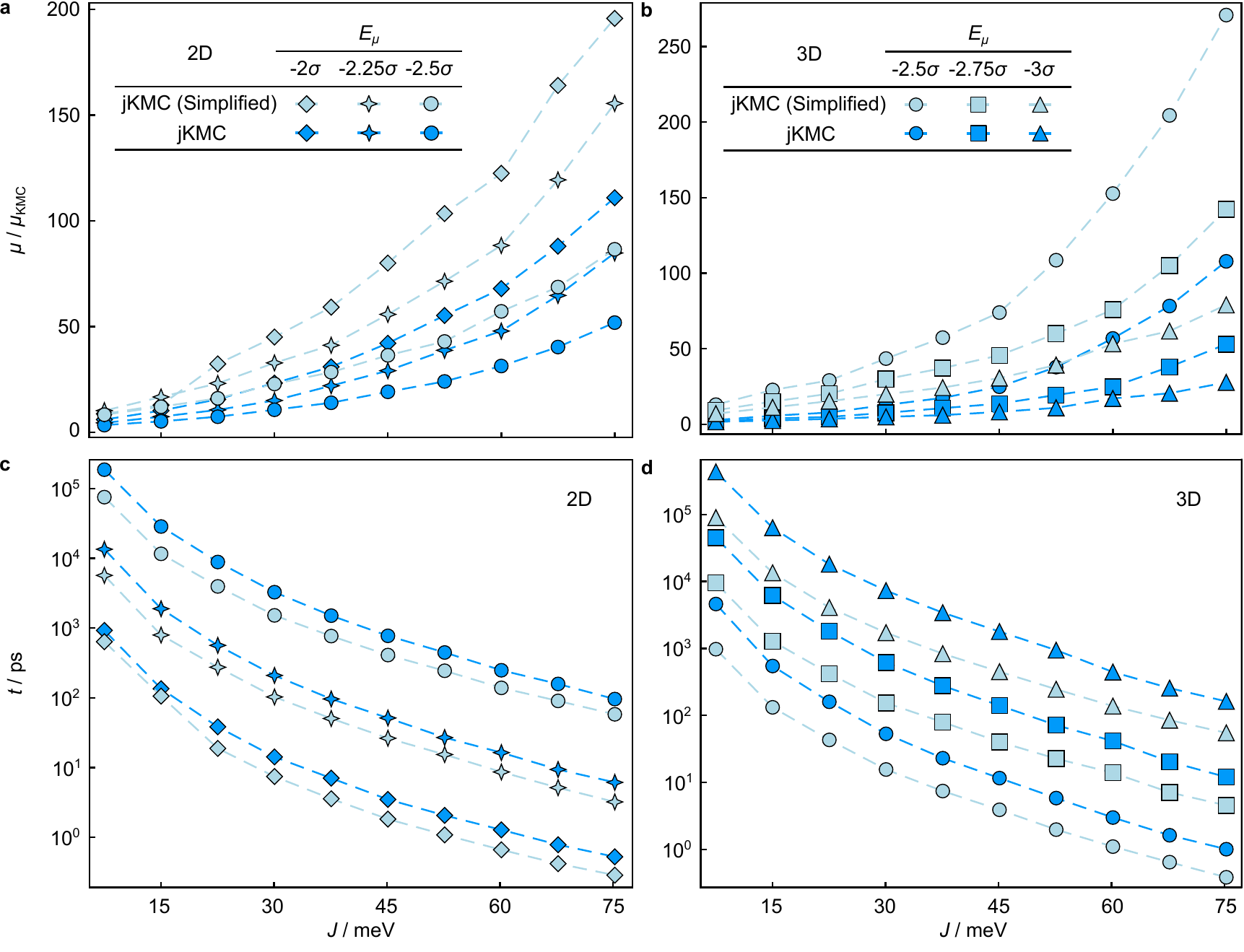}
    \caption{\textbf{Simplified jKMC produces mobilities on the same order of magnitude as jKMC.} 
    In both \textbf{(a)} two and \textbf{(b)} three dimensions, simplified jKMC reproduces jKMC mobilities to about a factor of 2 (for the parameters tested), but tends to systematically overestimate them. Results are calculated for $\sigma=\SI{150}{meV}$, $\lambda=\SI{200}{meV}$, and $T=\SI{300}{K}$.
    \textbf{(c,d)} The corresponding times taken to reach the target energy $E_\mu$ using simplified jKMC.}
    \label{fig:Results_S}
\end{figure*}

The mobility enhancements and times taken to reach the target energy $E_\mu$ for simplified jKMC are shown in \cref{fig:Results_S}.

\section{Effective IPR convergence}
\label{app:Effective_IPR_convergence}
\begin{figure*}
    \centering
    \includegraphics[width=\textwidth]{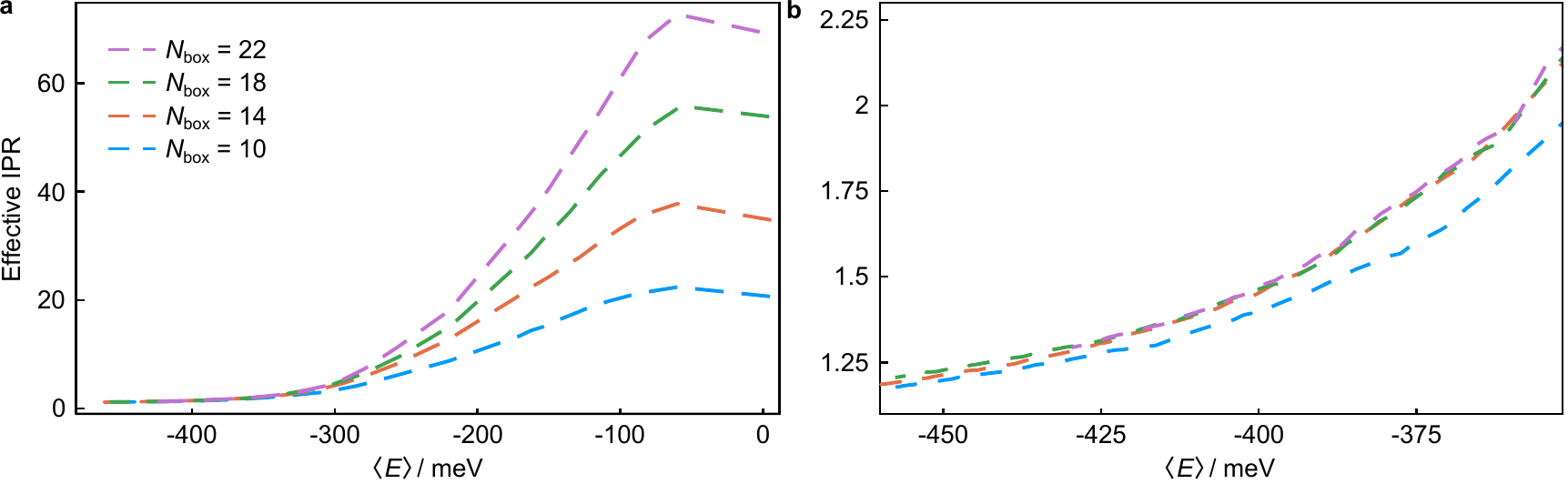}
    \caption{\textbf{Calculations of effective IPRs at deep $\langle E \rangle$ converge rapidly as a function of lattice size.} The effective IPR in $d$ dimensions is calculated using the neighbourhood-averaging method on a lattice of size $N_\mathrm{box}^d$. 
    \textbf{(a)} Calculations of the effective IPR converge when $N_\mathrm{box}$ is large enough to accommodate the sizes of the delocalised polaron states that contribute to transport at $\langle E \rangle$. Here, $N_\mathrm{box}\approx 20$ is sufficient for convergence for $\langle E\rangle$ below about \SI{-300}{meV}. By contrast, dKMC requires boxes large enough to converge the mean IPR of all states, corresponding to $\langle E \rangle=0$, which would require substantially larger $N_\mathrm{box}$.
    \textbf{(b)} Enlarged view of panel \textbf{a} for the values of $\langle E \rangle$ considered in this paper. Here, smaller $N_\mathrm{box}$ suffice to achieve convergence because the effective IPR is a thermal average, mostly sensitive to low-energy, localised states that are adequately described using small lattices.  Results are shown for three dimensions, $J=\SI{75}{meV}$, $\sigma=\SI{150}{meV}$, $\lambda=\SI{200}{meV}$, and $T=\SI{300}{K}$. }
    \label{fig:IPR_E_convergence}
\end{figure*}

\Cref{fig:IPR_E_convergence} shows that the lattices diagonalised in jKMC can  be much smaller than those in dKMC. Smaller lattices are much easier to diagonalise, giving jKMC a significant computational advantage over dKMC.

\end{document}